\documentclass[onecolumn]{aastex63}

\newcommand\kak[1]{\left(#1\right)}
\newcommand\dfrac[2]{\displaystyle \frac{#1}{#2}}

\accepted{January 21, 2022}
\submitjournal{Astrobiology, https://doi.org/10.1089/ast.2021.0126}

\shorttitle{Inefficient water degassing inhibits ocean formation on rocky planets}
\shortauthors{Miyazaki and Korenaga}
\watermark{}

\begin{document}

\title{Inefficient water degassing inhibits ocean formation on rocky planets:\\ An insight from self-consistent mantle degassing models}

\correspondingauthor{Yoshinori Miyazaki}
\email{ymiya@caltech.edu}

\author{Yoshinori Miyazaki} 
\affiliation{Department of Earth and Planetary Sciences, Yale University \\
New Haven, CT 06511, USA}
\affiliation{Division of Geological and Planetary Sciences, California Institute of Technology\\
 Pasadena, CA 91125, USA}

\author{Jun Korenaga} 
\affiliation{Department of Earth and Planetary Sciences, Yale University \\
New Haven, CT 06511, USA}


\begin{abstract}
A sufficient amount of water is required at the surface to develop water oceans. A significant fraction of water, however, remains in the mantle during magma ocean solidification, and thus the existence of water oceans is not guaranteed even for exoplanets located in the habitable zone. To discuss the likelihood of ocean formation, we built two models to predict the rate of mantle degassing during the magma ocean stage and the subsequent solid-state convection stage. We find that planets with low H$_2$O/CO$_2$ ratios would not have a sufficient amount of surface water to develop water oceans immediately after magma ocean solidification, and the majority of the water inventory would be retained in the mantle during their subsequent evolution regardless of planetary size. This is because oceanless planets are likely to operate under stagnant lid convection, and for such planets, dehydration stiffening of the depleted lithospheric mantle would limit the rate of mantle degassing. In contrast, a significant fraction of CO$_2$ would already be degassed during magma ocean solidification. With a strong greenhouse effect, all surface water would exist as vapor, and water oceans may be absent throughout planetary evolution. Volatile concentrations in the bulk silicate Earth are close to the threshold amount for ocean formation, so if Venus shared similar concentrations, small differences in solar radiation may explain the divergent evolutionary paths of Earth and Venus.
\end{abstract}




\section{Introduction} \label{sec:intro}
The continuing discovery of extrasolar terrestrial planets has increased the prospect of identifying habitable exoplanets. Earth-size exoplanets are starting to be found at a heliocentric distance where liquid water may be stable \citep{Gillon2017, Gilbert2020}, and one of the super-Earth size exoplanets identified in the habitable zone, K2-18b, has water vapor detected in its atmosphere \citep{Benneke2019b, Tsiaras2019}. With the upcoming James Webb Space Telescope, more detections of water vapor are expected on potentially habitable exoplanets \citep{Gardner2006}, and the possible presence of surface water makes such exoplanets a leading candidate to search for biosignatures \citep{Schulze-Makuch2011, Madhusudhan2020}. Given the importance of surface liquid water, one of the key questions is the likelihood of water ocean formation on an exoplanet with a water vapor atmosphere. 

The presence of surface water has often been discussed in terms of the habitable zone \citep{Kasting1993, Kopparapu2013, Kodama2019}, which is the region where liquid water is stable at the temperature and pressure conditions of a planetary surface. The premise, however, is that the amount of surface water is large enough to develop oceans. Although a threshold value to form water oceans is $\sim 10^{-5}$ ocean mass (one ocean mass = $1.4\times10^{21}$~kg) for the present-day Earth and thus well exceeds the current ocean mass, the threshold changes with the amount of greenhouse gases and net stellar radiation \citep{Abe1993a, Salvador2017}. For example, during the early Hadean on Earth, an atmosphere of 100~bar CO$_2$ would have increased this threshold to $\sim$0.1 ocean mass, so it is not guaranteed that the threshold amount of water would always exist on the surface of terrestrial planets. A significant fraction of the water inventory can be retained in the planetary interior, precluding the formation of water oceans. 

The initial amount of surface water is determined by volatile partitioning during magma ocean solidification, whereas its subsequent evolution is governed by interaction with mantle convection \citep{Ito1983, Korenaga2017b}. Previous studies have often assumed that the mantle entirely degases during magma ocean solidification \citep{Elkins-tanton2008, LeBrun2013, Salvador2017}, but considering the rheological transition of a partially molten medium and its slow compaction velocity, a significant amount of volatiles could be trapped in the mantle \citep{Hier-Majumder2017, Miyazaki2022a}. Furthermore, if water oceans are initially absent, the lack of surface water may have persisted for a geological timescale. The presence of surface water is critical in reducing the lithospheric strength and thus in triggering plate tectonics \citep{Korenaga2007, Korenaga2020}. Without oceans, the mantle may be in the mode of stagnant lid convection, and mantle upwelling would be suppressed by a rigid lid covering the entire surface \citep{Solomatov1995}. Furthermore, dehydration stiffening of the depleted lithospheric mantle, caused by loss of water upon melting, could further hinder mantle processing. Under such a mode of convection, mantle degassing becomes inefficient, and the amount of surface water could be limited over a substantial period of time.



In this paper, we solve for mantle degassing during and after the solidification of a magma ocean to investigate the likelihood of ocean formation on terrestrial planets. Initial conditions necessary for planets to develop water oceans are explored for various sizes and volatile compositions. Previous studies discussing the impact of mantle degassing on habitability have mostly focused on the solid-state convection stage \citep{Noack2017, Tosi2017, Vilella2017, Dorn2018}, but the atmosphere could be dominated by volatiles degassed during the preceding magma ocean. Moreover, volatile concentrations in the mantle affect viscosity and thus the thermal evolution of planets, so it is crucial to solve for the entire history of mantle degassing self-consistently. 
It is noted that we assume that planets are in an oxidized condition, and the atmospheric components considered here are limited to H$_2$O and CO$_2$. The redox of the early Earth remains controversial, but recent studies suggest that planets larger than Mars-size would have an oxidized atmosphere \citep{Hirschmann2012, Deng2020}.

\section{Degassing during magma ocean: Prior to the solidification of the mantle surface} \label{sec:mo}
We first consider mantle degassing during the early stage of magma ocean solidification. Here the focus is on the aftermath of the final magma ocean produced by giant impacts, which are common in $N$-body simulations of planetary formation \citep[e.g.,][]{Rubie2015, Quintana2016}. The final one was likely energetic enough to melt most of the mantle \citep{Nakajima2015}, so the atmosphere existing prior to the impact would either have been lost \citep{Genda2005} or have re-equilibrated with the magma ocean. The same is likely to be true for larger exoplanets because, with an impactor of 10\% target mass, a typical impact velocity is high enough to produce substantial melting of the mantle \citep{Stixrude2014}. Also, the amount of radiogenic Xe in the atmosphere suggests that a catastrophic atmospheric loss likely occurred around the time of the Moon-forming giant impact \citep{Porcelli2000}. Therefore, the final atmospheric mass or composition would not be affected by mantle degassing during earlier magma oceans, and we estimate how the volatile budget after the final giant impact would be distributed between the atmosphere and the mantle.

\begin{figure}
\centering
\includegraphics[width=0.8\textwidth]{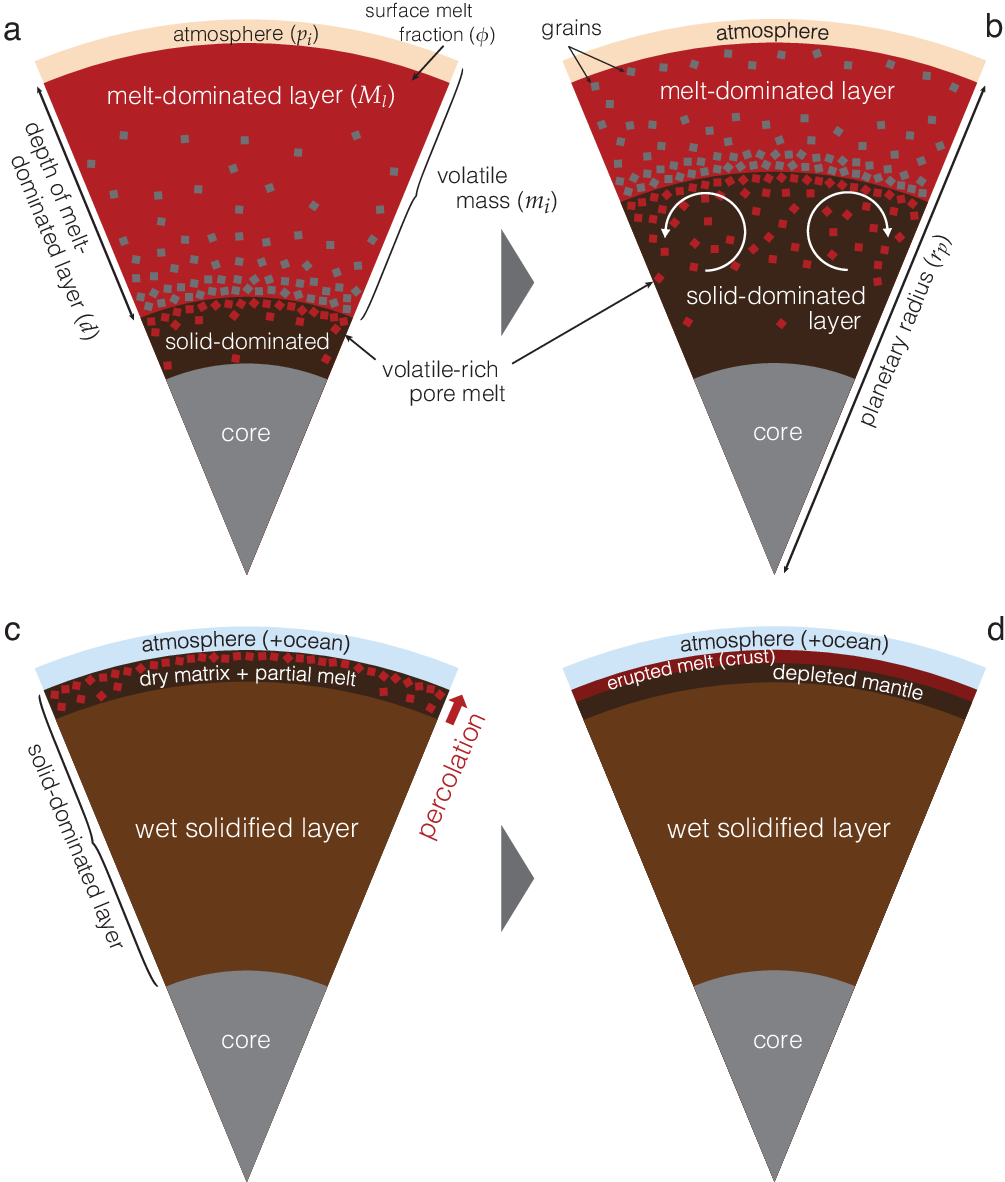}
\caption{Schematic illustration of the upper mantle and atmosphere during the solidification of magma ocean. (a) Magma ocean undergoes the rheological transition from the bottom, forming a solid-dominated layer. Melt would be trapped in pore spaces with volatiles, and thus volatiles would not concentrate in the melt-dominated layer. Variables used in Equation~(\ref{massc}) are labeled in corresponding reservoirs. (b) Magma ocean continues to solidify, and the thickness of the solid-dominated layer grows. Rapid convection and the Rayleigh-Taylor instability would efficiently cool the melt- and solid-dominated layers, respectively, both creating an adiabatic thermal profile. Because of the rapid onset of the instability, volatiles would efficiently be sequestered in the deeper region. 
(c) When the melt-dominated layer disappears, convective heat flux plummets, and the surface temperature drops below the mantle solidus. When sufficient amount of water is degassed, oceans may form. Because melt percolation is faster than solid-state convection, percolation contributes to additional degassing. (d) Erupted melt forms a crust at the top, leaving a dry depleted layer below the crust. Because the deeper region is completely solidified, percolation would not occur, and the most of the mantle would remain hydrated.  }
\label{fig_cartoon}
\end{figure}

The solidification of a magma ocean can be divided into two stages by its surface melt fraction (Figure~\ref{fig_cartoon}). With a high melt fraction at the surface, the magma ocean produces a high convective heat flux, resulting in a surface temperature high enough to maintain a molten surface \citep{LeBrun2013, Salvador2017}. As the mantle cools down and melt fraction becomes lower than the critical value ($\sim$0.4), however, the partially molten mantle would undergo a rheological transition and start to behave as solid \citep{Abe1993b}. Convective heat flux plummets, and the surface temperature would drop to $<$500~K, which is well below the mantle solidus temperature \citep{LeBrun2013, Miyazaki2022a}. Therefore, the surface layer would be completely solidified in a short timescale once the surface melt fraction reaches the critical value, even though the interior may still be partially molten. Chemical equilibrium between the mantle and the hydrosphere is not maintained thereafter, and volatile exchange between the surface and the interior would instead be characterized by solid-state convection \citep{Abe1997}. 

In this section, we focus on volatile partitioning between the atmosphere and magma ocean while a partially molten surface is sustained, and the main goal here is to show the impact of the rheological transition on mantle degassing. We emphasize that the magma ocean considered here is produced by the final giant impact on the planet. Other degassing models have also examined the same period \citep[e.g.,][]{Elkins-tanton2008, Lebrun2013, Salvador2017, Bower2019}, and ours and these studies investigate a different evolutionary stage from accretionary magma ocean models \citep[e.g.][]{Matsui1986, Zahnle1988}. The latter type of model tracks the amount of added surface water as an impactor hits the planet and considers different degassing processes from the former. Also, accretionary magma ocean models do not consider large-scale melting at the final stage of accretion, so the mantle and atmosphere are never in equilibrium in terms of volatile partitioning. We assume that a giant impact most likely induces global-scale melting \citep{Nakajima2015} so that a substantial amount of volatiles would be partitioned into the mantle. Although both types of models aim to quantify surface water mass at the end of accretion, we caution that their validity depends on the assumed scale of mantle melting.

\subsection{Volatile partitioning} \label{sec:partition}
When a magma ocean behaves rheologically as liquid, the atmosphere and the magma ocean are equilibrated in a short timescale, and volatile partitioning between the magma and the atmosphere would determine the atmospheric pressure \citep{Elkins-tanton2008, LeBrun2013}. Magma at the surface dissolves volatiles up to their solubility limits, and volatiles in excess of saturation would reside in the atmosphere. We solve for volatile partitioning and mass balance at each time step as the magma ocean undergoes the rheological transition from the bottom upward \citep[e.g.,][]{Abe1993b, Solomatov2007}. 

Volatile concentration in the surface magma, $x_i$, corresponds to the solubility at its atmospheric partial pressure, $P_i$, for each volatile species $i$. The solubility measurements of CO$_2$ and H$_2$O are parameterized as follows \citep{Blank1994, Lichtenberg2021a}:
\begin{equation}
	P_\mathrm{H_2O} = \kak{\frac{x_\mathrm{H_2O}}{6.8 \times 10^{-8}}}^{1.42},
\end{equation}
\begin{equation}
	P_\mathrm{CO_2} = \kak{\frac{x_\mathrm{CO_2}}{4.4 \times 10^{-12}}},
\end{equation}
where pressure is shown in the unit of Pa. Some models indicate higher solubility for both H$_2$O and CO$_2$ \citep{Papale1997, Gardner1999}, but we adopt the least soluble model to prepare the most optimistic case for ocean formation.  
The variables $P_i$ and $x_i$ can be solved using the following mass balance \citep{Bower2019}:
\begin{equation} \label{massc}
	\phi x_i M_l + \frac{\mu_i}{\overline{\mu}} \frac{4 \pi r_p^2}{g} P_i = m_i,
\end{equation}
where $\phi$ is the melt fraction at the surface, $M_l$ is the mass of the melt-dominated layer, $\mu_i$ is the molar mass of volatile $i$, $\overline{\mu}$ is the mean molar mass of the atmosphere, $r_p$ is the planetary radius, and $m_i$ is the total mass of volatile $i$ in the atmosphere and the melt-dominated layer (Figure~\ref{fig_cartoon}a). The surface melt fraction $\phi$ is calculated to be consistent with the depth of the melt-dominated layer \citep{Miyazaki2019b}, considering that the magma ocean would have an adiabatic temperature profile as a result of rapid convection.

The volatile mass $m_i$ represents those that were not trapped in the solid-dominated layer existing in the deeper region. Taking into account the rheological transition of a partially molten medium, a significant amount of melt could be trapped in the pore space of the solid matrix, together with volatiles dissolved in the melt (Figure~\ref{fig_cartoon}b). The percolation of pore melt is slower than the speed of magma ocean solidification \citep{Hier-Majumder2017}, and a newly formed rheologically-solid layer, which includes volatile-rich pore melt, would be quickly delivered to the deeper region by the Rayleigh-Taylor instability \citep{Maurice2017, Miyazaki2019b}. For parameters used in \cite{Miyazaki2019b}, the entire lower mantle solidifies in $\sim$4000~years, whereas the timescale for the Rayleigh-Taylor instability is less than a year just after the lower mantle undergoes the rheological transition. Volatiles trapped in the pore space of the solid matrix would thus be effectively sequestered in the deep mantle and be segregated from the remaining melt layer (Figure~\ref{fig_cartoon}b).

During the instability-triggered downwelling, the melt in the pore space would solidify by adiabatic compression, and the melt would be supersaturated and exsolve volatiles. Yet, the saturation limit would be reached only when melt fraction becomes lower than $\sim$0.002. Here we consider an initial H$_2$O content below 0.1~wt\%, and in such a case, concentrating volatiles by 500 times would still not result in exceeding saturation limits \citep{Kawamoto1997}. Therefore, the melt-solid mixture would be mostly solidified when volatiles in melt pockets finally start to be degassed, and exsolved volatiles would either diffuse into the surrounding solid phase to form nominally anhydrous minerals or be trapped in the solid matrix as bubbles. We thus assume that volatiles incorporated in the pore space of the solid matrix are kept in the solid-dominated layer, and the corresponding amount is subtracted from the volatiles budget of the melt-dominated layer $m_i$ at each time step.

\subsection{Degassing of the upper mantle by percolation} \label{sec:perc}
\begin{figure}
\includegraphics[width=1\textwidth]{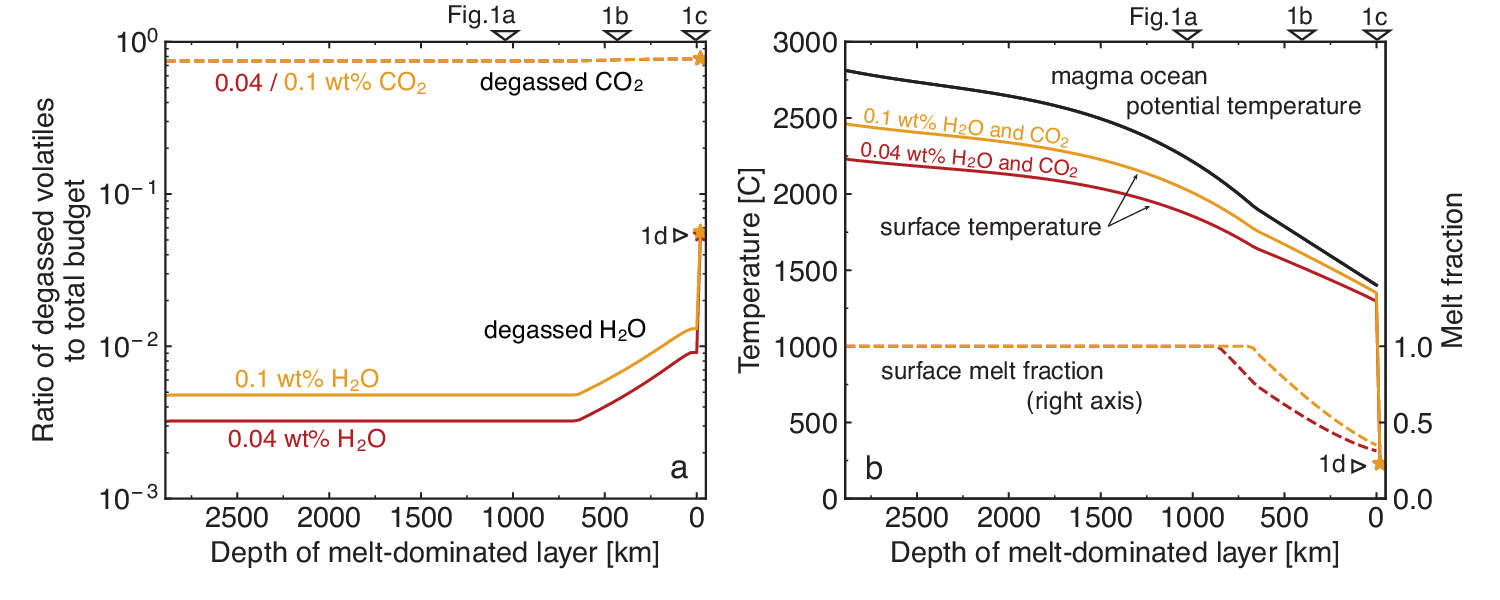}
\caption{The evolution of (a) volatiles in the atmosphere and (b) the thermal state as a function of the melt-dominated layer depth. As solidification proceeds, the layer thickness decreases, so the system evolves from left to right. (a) The ratio of degassed volatiles to the total volatiles budget for H$_2$O (solid) and CO$_2$ (dashed). Initial mantle concentrations of 0.04~wt\% (red) and 0.1~wt\% (yellow) are tested for both volatiles. For H$_2$O, the two values correspond to 1.2 and 3~ocean masses, respectively. (b) The mantle potential temperature (black solid) and the surface temperature (colored solid, left axis) with the melt fraction of magma ocean at the surface (dashed, right axis). The snapshots of the mantle structure are shown in Figure~\ref{fig_cartoon}, with triangles indicating the corresponding panels. }
\label{fig_degas}
\end{figure}

Once the entire mantle undergoes the rheological transition and the melt-dominated layer disappears, the evolution of a magma ocean is characterized by solid-state convection. Although percolation is slower than the Rayleigh-Taylor instability, it is faster than solid-state convection \citep{Miyazaki2022a}, so melt and dissolved volatiles in the pore space would escape towards the surface (Figure~\ref{fig_cartoon}c). Volatiles included in the partial melt would be degassed to the hydrosphere, and their amount is proportional to the volume of partially molten layer when the melt-dominated layer disappears. We calculate its thickness from the adiabatic temperature profile with a critical melt fraction of 0.4 at the surface (see Appendix~\ref{sec:degas}) \citep{Abe1993b, Solomatov2007}. For simplicity, the mantle is assumed to be compositionally homogeneous and have a pyrolitic composition, and in such a case, the partial melt layer would extend to a depth of ~5 GPa (Appendix A.3). We note that, if the mantle experiences differentiation during magma ocean solidification, the thickness of the partial melt layer could decrease by half \citep{Miyazaki2022a}, and degassing by percolation could be smaller than presented in Figures \ref{fig_degas}-4. The mantle, however, would eventually be homogenized, and the long-term evolution of surface water would converge to results presented in Section 3.

\subsection{Results} \label{sec:res2}
Figure~\ref{fig_degas} shows a typical evolution of the atmospheric pressure for an Earth-size planet at 1~AU. A fraction of volatiles would be degassed from the beginning to maintain equilibrium with volatiles in the mantle, but further degassing is limited during the early stage of solidification because volatiles are trapped in the pore space of the solid-dominated layer and thus volatile concentrations remain mostly unchanged (Figure~\ref{fig_degas}a). Here we assume initial volatile concentrations of 0.04 and 0.1~wt\% for both H$_2$O and CO$_2$, and regardless of the initial concentration, $>$99\% of the total H$_2$O budget is estimated to be retained in the mantle during the early stage of magma ocean solidification. On the other hand, because CO$_2$ is less soluble to silicate magma, $\sim$70\% of the total CO$_2$ would be released to the atmosphere \citep[Figure~\ref{fig_degas}a;][]{Hier-Majumder2017}, and such an atmosphere dominated by CO$_2$ would further suppress the degassing of lighter gases, including H$_2$O \citep{Bower2019}. Degassing resumes when the surface magma starts to solidify as volatiles concentrate in the remaining melt phase (Figure~\ref{fig_degas}b) and a higher atmospheric pressure is required to maintain equilibrium between the atmosphere and the magma ocean. Degassing by percolation would also increase the volatile mass in the atmosphere, yet $\sim$95\% of H$_2$O would still be retained in the mantle.

\begin{figure}
\centering
	\begin{minipage}[l]{0.5\textwidth}
	\includegraphics[width=0.9\linewidth]{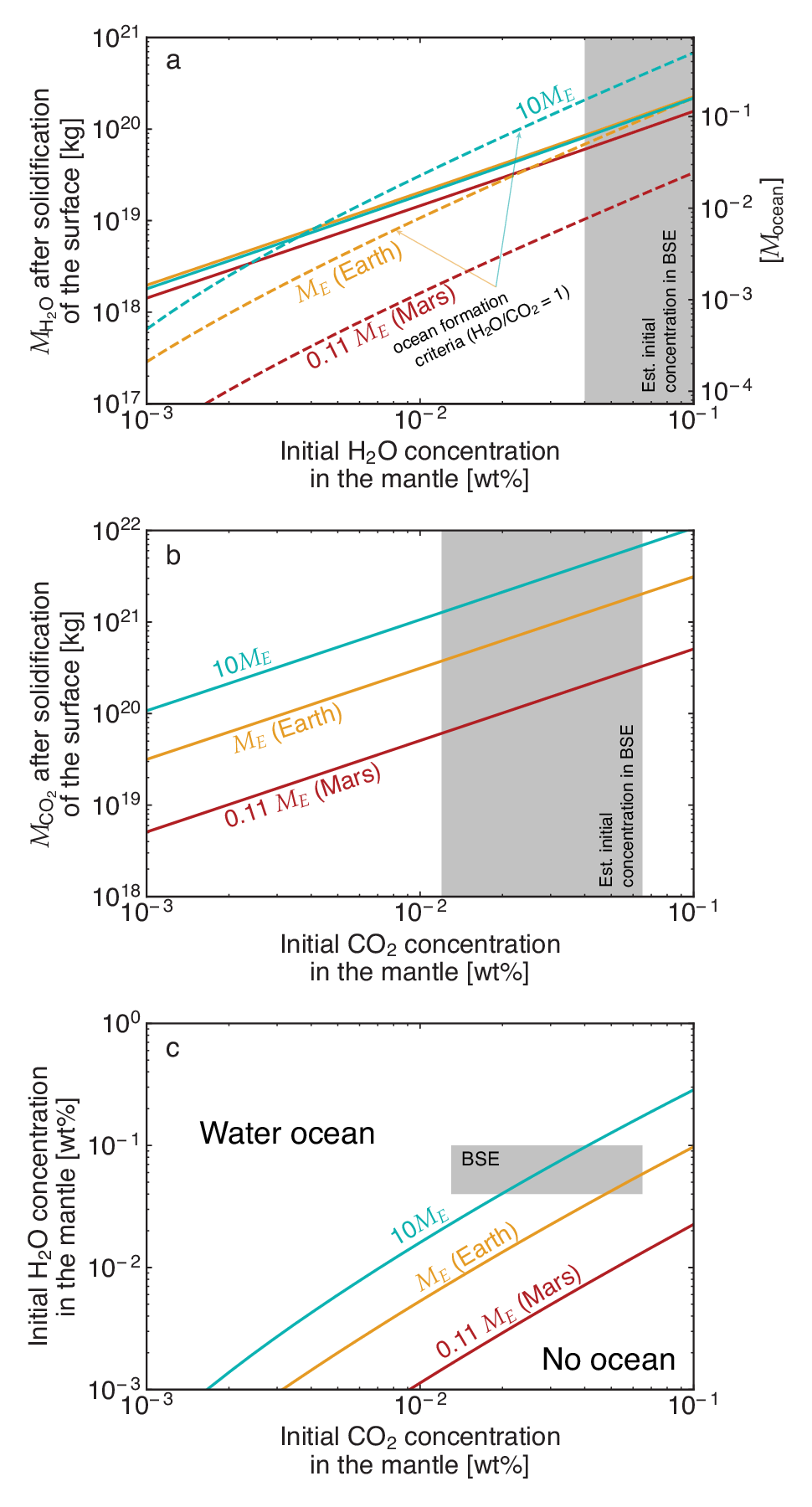}
	\end{minipage}
	\begin{minipage}[c]{0.47\textwidth}
	\caption{(a, b) Atmospheric pressures of (a) H$_2$O and (b) CO$_2$ when the surface of the magma ocean solidified, as a function of the initial volatile concentration. The pressure can be translated to the total amount of degassed volatiles through Equation~(\ref{p2M}), and the pressure is calculated assuming that all volatiles exist as gas. Colors denote different planetary sizes: Mars- (red), Earth-size ($M_E$, yellow), and 10$M_E$ super-Earths (cyan). The value of $p_\mathrm{H_2O}$ is similar among planets of different size sizes, and all lines are very close to each other. In (a), the minimum amount of H$_2$O necessary to stabilize water oceans is also plotted, assuming that the H$_2$O/CO$_2$ mass ratio is 1. Mars- and Earth-size planets would have sufficient amount of water to form oceans, whereas super-Earths would lack oceans immediately after the solidification of the mantle surface. (c) Conditions necessary for water ocean formation as a function of initial H$_2$O and CO$_2$ concentrations. Colored lines separate the two regimes: the formation (upper left) and absence of water oceans (lower right) when the mantle surface solidified. With the same volatile concentrations, smaller planets are more likely to develop water oceans. Shaded rectangles represent the estimated volatile concentrations in the bulk silicate Earth (BSE) from \cite{Hirschmann2009} and \cite{Korenaga2017b}.}
	\label{fig_ocean}
	\end{minipage}
\end{figure}

We repeat this calculation for a range of initial volatile content and planetary mass. As expected, the pressure of degassed atmosphere increases with higher initial volatile concentrations, but a larger planetary mass does not necessarily result in a thicker atmosphere; the partial pressure for degassed H$_2$O atmosphere, $p_\mathrm{H_2O}$, shows little dependence on planetary mass (Figure~\ref{fig_ocean}a). This is because the majority of H$_2$O is stored in the mantle, and the water concentration in the mantle changes little from the initial value. Therefore, the corresponding atmospheric pressure in equilibrium with the mantle reservoir is mostly the same regardless of planetary mass. In contrast, most of the CO$_2$ inventory is degassed to the atmosphere, and thus $p_\mathrm{CO_2}$ becomes higher for larger planets (Figure~\ref{fig_ocean}b). The partial pressure is related to the mass of volatile $i$ in the atmosphere, $M_i$, through:
\begin{equation} \label{p2M}
	p_i = \frac{M_i g}{4 \pi r_p^2} \frac{\overline{\mu}}{\mu_i} \sim M_i,
\end{equation}
where $g$ is the gravitational acceleration and $r_p$ is the planetary radius, and the relation between $g$ and $r_p$ is calculated from the simplified interior model of \cite{Seager2007}. Therefore, with the same volatile concentrations, $M_\mathrm{CO_2}$ increases with planetary mass (Figure~\ref{fig_ocean}b), whereas $M_\mathrm{H_2O}$ remains the same for planets of any size (Figure~\ref{fig_ocean}a). 

When the amount of surface H$_2$O is small, the entire surface water budget can exist as vapor in the atmosphere without forming water oceans. The threshold amount of H$_2$O for ocean formation is estimated using a 1-D radiative-convective model of \cite{Nakajima1992}. The model predicts that the threshold amount increases with a thicker CO$_2$ atmosphere \citep{Abe1993, Salvador2017} because a thicker atmosphere can contain more H$_2$O vapor, and its amount further increases under a higher surface temperature induced by a stronger greenhouse effect. Under the same initial volatile concentrations, therefore, larger planets are less likely to develop water oceans immediately after the solidification of the mantle surface (Figure~\ref{fig_ocean}a, \ref{fig_ocean}c). For an initial CO$_2$ concentration of 0.01 wt\%, the amount of degassed CO$_2$ ($M_\mathrm{CO_2}$) increases from $5.1\times 10^{19}$~kg for a Mars-size planet to $1.1\times10^{21}$~kg for a 10$M_E$ super-Earth (Figure~\ref{fig_ocean}b), which would raise the amount of surface H$_2$O required to develop water oceans from ~$\sim 1.6\times10^{18}$~kg to $\sim 3.1 \times 10^{19}$~kg (Figures \ref{fig_ocean}a). Nevertheless, the amount of surface water ($M_\mathrm{H_2O}$) does not change with planetary size, where $M_\mathrm{H_2O}$ is $\sim 2 \times 10^{19}$~kg for any planets with an initial H$_2$O concentration of 0.01~wt\%. Smaller planets are thus more likely to develop water oceans when the surface of the mantle solidifies. How the initial volatile concentration of terrestrial planets is characterized is still debated \citep{Hirschmann2021, Li2021}, but if terrestrial exoplanets have similar volatile concentrations to Earth, super-Earths may lack water oceans when the surface of the mantle solidifies (Figure~\ref{fig_ocean}c).

It is noted that the minimum surface H$_2$O mass to form water oceans in Figure~\ref{fig_ocean}ca is associated with an uncertainty of a factor of $\sim$2 because a 1-D gray atmospheric model is adopted here. Considering that the subsiding branch of the atmospheric circulation is undersaturated with water vapor, the average relative humidity should be smaller than 1 \citep[e.g.,][]{Ishiwatari2002, Abe2011, Pierrehumbert2016}. It is thus assumed that the relative humidity is 0.7 throughout the troposphere in our model, but this value remains unconstrained for an atmosphere with an extreme CO$_2$ concentration and may range between 0.5 and 1. Also, CO$_2$ existing near the surface could be in a supercritical state when $p_\mathrm{CO_2}$ is larger than 74~bar. Although the temperature and pressure conditions considered here are distant from its critical point, physical properties including opacity may differ by some degree from the gas state. Implementing these factors, together with the wavelength dependence of the opacity, may modify our estimate of the minimum H$_2$O mass required for ocean formation by some factor. The overall trend of our results is unlikely to change, but a more detailed model would be needed to predict the actual threshold. 

\subsection{Discussion: early Earth and early Venus}
The amount of volatiles in the bulk silicate Earth (BSE) could have been just above the threshold for stabilizing water oceans at the point of magma ocean solidification. Although the volatile content in the terrestrial mantle remains controversial, recent estimates suggest that the mantle and the hydrosphere in total contain 1.2--3~ocean mass of H$_2$O \citep{Hirschmann2009, Korenaga2017b} and 4.8--26$\times$10$^{20}$~kg of CO$_2$, which translates to initial concentrations of 0.04--0.1~wt\% and 0.01--0.07~wt\%, respectively, in the mantle. The estimated volatile concentrations lie on the boundary between the regimes of ocean formation and dry surface (Figure~\ref{fig_ocean}c), and if BSE had a lower-bound H$_2$O and an upper-bound CO$_2$ concentration, water oceans may have been absent during the early stage of evolution. As we discuss in Section~3, the rate of mantle degassing decreases substantially after the mantle surface solidifies, and thus the surface would continue to lack water oceans if oceans are absent at the time of surface solidification. The presence of surface water in the Hadean has been suggested from zircon records \citep[e.g.,][]{Wilde2001, Mojzsis2001}, so water oceans were likely present on the surface of Earth from the beginning of its evolution. Therefore, a combination of a low H$_2$O and a high CO$_2$ concentration in BSE ($>$0.05~wt\%) may be ruled out if water oceans indeed existed on the surface of the Hadean Earth.

\begin{figure}
	\begin{minipage}[l]{0.5\textwidth}
	\includegraphics[width=\textwidth]{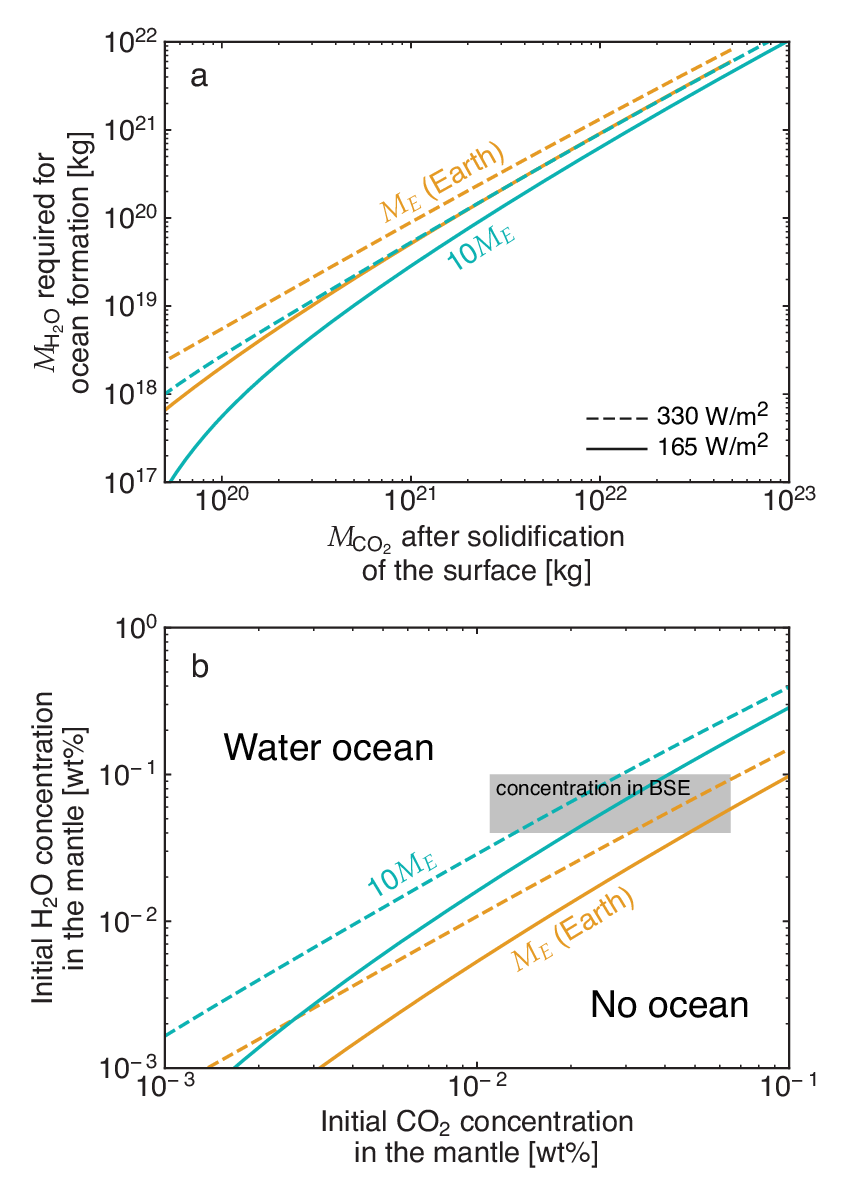}
	\end{minipage}
	\begin{minipage}[c]{0.47\textwidth}
	\caption{(a) The minimum amount of surface H$_2$O required to form water oceans as a function of $p_\mathrm{CO_2}$. The pressure can be converted to mass using Equation~(\ref{p2M}), and results are calculated for different levels of net stellar radiation: 165 (solid, early Earth) and 330~W~m$^{-2}$ (dashed, early Venus). Colors denote different planetary sizes: Earth-size ($M_E$, yellow) and 10$M_E$ super-Earth (cyan). (b) The same as Figure~\ref{fig_ocean}c but with different levels of net stellar radiation. Colored lines separate the two regimes: the formation (upper left) and absence of water oceans (lower right) when the planetary surface solidified. Planets further away from the central star are more likely to develop water oceans, but planetary size plays a larger role in characterizing ocean formation. Shaded rectangles represent the estimated volatile concentrations in the bulk silicate Earth (BSE) from \cite{Hirschmann2009} and \cite{Korenaga2017b}. }
		\label{fig_Venus}
		\end{minipage}
\end{figure}

The early Venus, on the other hand, may have lacked water oceans because a larger amount of surface H$_2$O is required for planets closer to the central star (Figure~\ref{fig_Venus}a). Assuming the same albedo, the net solar radiation for the early Venus was 330~W~m$^{-2}$, which is two times higher than the value for the early Earth. To balance stronger solar radiation, the overall thermal structure, including the planetary surface, becomes hotter, thus allowing more H$_2$O to exist as water vapor in the atmosphere \citep{Nakajima1992}.  The minimum amount of H$_2$O to form oceans doubles for a $p_\mathrm{CO_2}$ of 100~bar and increases by an order of magnitude for $p_\mathrm{CO_2}$$<$10~bar. As a result, for a planet receiving 330~W~m$^{-2}$ of stellar radiation, water oceans would be absent for a wider range of initial volatile concentrations. If Venus had similar volatile concentrations to Earth, our result suggests that Venus would develop water oceans only when the total amount of CO$_2$ is less than $9.7 \times 10^{20}$~kg and the mantle is sufficiently wet (Figure~\ref{fig_Venus}b). Because the threshold for ocean formation lies within the estimated volatile concentrations in BSE, the location of planets and thus the amount of net solar radiation may have decided the fate of two similar planets (see Section~\ref{sec:dis3} for further discussions). Mars, on the other hand, likely developed water oceans immediately after magma ocean solidification because of its small size (Figure~\ref{fig_ocean}c) and its distance from the Sun (Figure~\ref{fig_Venus}b).

Volatile concentrations required to develop water oceans is much higher in our model than estimated in \cite{Salvador2017} because we take into account volatiles that are trapped in the pore space of the melt-solid mixture. Models of magma ocean solidification have often neglected this effect and assumed that volatiles, behaving as incompatible elements, concentrate in the residual melt layer as the magma ocean solidifies from the bottom to the top \citep{Elkins-tanton2008, Hamano2013, Lebrun2013, Salvador2017}. In such a scenario, the concentration of volatiles eventually exceeds saturation levels and results in efficient degassing. Volatiles, however, would not be efficiently degassed because the compaction of melt-solid mixture is slow \citep{Hier-Majumder2017} and because the Rayleigh-Taylor instability efficiently delivers volatile-rich pore melt to the deeper mantle. Therefore, the remaining melt layer near the surface experiences little change in volatile concentrations, and the mantle is likely to remain hydrated (Section~\ref{sec:partition}, Figure~\ref{fig_degas}a).

\section{Degassing during solid-state convection:\\ After the solidification of the mantle surface}
Once the surface of the mantle solidifies, mantle degassing during the subsequent evolution would be characterized by solid-state convection \citep{Miyazaki2022a}. Although partial melt may still exist in the interior, the planetary surface would be covered by a solidfied lid, and the mode of volatile exchange would be similar to what is seen on the present-day Venus and Mars \citep{Noack2017, Dorn2018}. The solid lid prohibits the mantle and the hydrosphere to reach equilibrium, and instead the evolution of the atmosphere becomes controlled by a balance between ingassing to and degassing from the mantle. The rate of ingassing and degassing differs appreciably depending on the mode of solid-state convection, so whether the mantle operates under plate tectonics or stagnant lid convection would control the atmospheric evolution of terrestrial planets. Here, plate tectonics refers to a mode that allows the continuous recycling of the uppermost layer, whereas in stagnant lid convection, a rigid immobile shell covers the surface, and convective motion is limited to the interior region beneath the shell \citep{Solomatov1995}. 

The key to triggering plate tectonics is the weakening of oceanic lithosphere \citep{Moresi1998}, and several mechanisms have been proposed, including grain-size reduction \citep[e.g.,][]{Kameyama1997, Bercovici2012} and thermal cracking \citep{Korenaga2007}. Comparison between different mechanisms are discussed in several reviews \citep{Bercovici2015, Korenaga2020}, but the weakening of the strongest part of the lithosphere requires the reduction of friction coefficient by some means, for which thermal cracking is so far the only mechanism that is consistent with our understanding of rock mechanics. Positive feedback between thermal cracking and serpentinization could potentially hydrate the lithosphere deeply \citep{Korenaga2017a}, with trapped water reducing lithostatic-hydrostatic pressure difference and thus effective friction coefficient. The presence of surface water is likely the key to lowering the yield strength of the lithosphere, and although the mechanism is different, the importance of surface water in triggering plate tectonics has been suggested in other studies as well \citep{Regenauer-Lieb2001, Gerya2008}. Indeed, other oceanless planets, including Venus, Mars, and Moon, are generally considered to be operating under the mode of stagnant lid convection. 

The absence of surface water thus leads to stagnant lid convection, which limits both ingassing and outgassing compared to plate tectonics \citep{Kite2009}. Under solid-state convection, the degassing of the mantle is triggered when the mantle material undergoes partial melting during its upwelling \citep{Fraeman2010, Orourke2012}, but a rigid immobile shell would limit the extent of such upwelling motion. With the same convective velocity, the processing of the mantle would be significantly slower for stagnant lid convection than plate tectonics, under which the mantle material is brought up all the way to the surface. Oceanless planets would therefore have a smaller supply of H$_2$O from the mantle than those with oceans, implying that an oceanless state may persist for the geological time.

The lack of surface recycling also prevents the transport of volatiles from the hydrosphere to the mantle, whereas the sequestration of volatiles is possible under plate tectonics \citep{Sleep2001a, Miyazaki2022a}. Volatiles, in particular CO$_2$, can be stored in oceanic crust as carbonate minerals \citep[e.g.,][]{Alt1999}, which could then be delivered to the interior by subduction. The majority of atmospheric CO$_2$ existed after magma ocean solidification on Earth should have been transported to the mantle by the end of Hadean \citep{Catling2020}, and a rapid removal of atmospheric CO$_2$ is also expected for other terrestrial planets with plate tectonics. On the other hand, the removal of atmospheric CO$_2$ would be prohibited for stagnant lid convection, and rather CO$_2$ would continue to accumulate in the atmosphere by mantle degassing. The threshold amount of H$_2$O to develop water oceans thus continues to increase with time for such planets, which would be another factor that further delays the formation of water oceans. 

Here, we estimate how long it takes for initially oceanless dry planets to develop water oceans after solid-state convection starts, by modeling the thermal evolution of terrestrial planets. Previous studies discussing habitability in the context of stagnant lid convection have focused on the degassing of CO$_2$ because the availability of greenhouse gas on the surface was considered to regulate the outer edge of the habitable zone \citep{Noack2017, Tosi2017, Vilella2017, Dorn2018}. However, they have not taken into account the massive CO$_2$ atmosphere released during magma ocean (Figure~\ref{fig_degas}a), and with its effect included, the amount of greenhouse effect should not be a limiting factor for habitability for planets with stagnant lid convection. Because the focus of this paper is the formation of water oceans on terrestrial planets, we do not model the evolution after water oceans form, but its impact on habitability is discussed in Section~\ref{sec:hab}.


\subsection{Methods: Mantle degassing under stagnant lid convection} \label{sec:met3}
The degassing rate of terrestrial planets that initially lack water oceans is calculated based on a heat flow scaling of stagnant lid convection \citep{Korenaga2009}. This scaling has been applied to the evolution of Mars, Venus, and super-Venus planets \citep{Fraeman2010, Orourke2012}, where they modeled the thermal evolution of crust, mantle, and core self-consistently by incorporating the influence of mantle processing on rheology and radiogenic heating. Our theoretical formulation follows \cite{Fraeman2010} and \cite{Orourke2012}, and readers are referred to these studies for details. We summarize the key equations of our model in the following, focusing on the difference from previous models. 

Our primary goal is to estimate the processing rate of the mantle to track the atmospheric evolution over time: melting by adiabatic decompression generates new crust and depleted mantle lithosphere (DML), and volatiles are degassed during crust formation.  A wide range of initial conditions were explored in \cite{Orourke2012}, but sub-solidus convection likely starts immediately after the surface of the mantle solidifies and degassing by percolation takes place (Section~\ref{sec:perc}). We thus adopt the thermal structure and volatile concentrations predicted in Section~\ref{sec:mo} as initial conditions to model the history of mantle degassing in a self-consistent manner.

\subsubsection{Scaling of stagnant lid convection} \label{sec:Nu}
The vigor of convection is characterized by the Nusselt number: convective heat flux normalized by conductive heat flux. The Nusselt number is calculated through a local stability analysis of the top thermal boundary layer \citep{Korenaga2009}, and we consider a condition where a bottom fraction of the boundary layer becomes gravitationally unstable. The Nusselt number depends on the following variables; the mantle potential temperature, $T_m$, the temperature difference across the thermal boundary layer, $\Delta T$, the mantle viscosity, $\eta_m$, viscosity increase by dehydration, $\Delta \eta_w$, and the thickness of the depleted lithospheric mantle, $h_\mathrm{DML}$. The temperatures $T_m$ and $\Delta T$ are solved from the energy balance of the mantle, and $\eta_m$ is calculated as a function of potential temperature $T_m$ and volatile concentrations in the mantle (Section~\ref{sec:rheo}). The evolution of $h_\mathrm{DML}$ is controlled by the combination of mantle processing, delamination, and rewetting (Section~\ref{sec:proc}). 

DML lacks volatiles, so the layer is more viscous than the source mantle. Thermal convection is suppressed with such effect of dehydration stiffening \citep{vanThienen2007}, and its effect is stronger for a higher mantle potential temperature because of a thicker DML. For a potential temperature of 1600~$^\circ$C, the thermal boundary layer becomes thicker by a factor of $\sim$5 than what is predicted in a scaling without dehydration \citep{Korenaga2009}. Therefore, hotter mantle does not indicate more efficient thermal convection, and considering the effect of dehydration stiffening is crucial for estimating $Nu$. The thickness of the thermal boundary layer (TBL) can be calculated as $h_\mathrm{TBL} = h_m/Nu$, where $h_m$ is the depth of the convective mantle.

\subsubsection{Mantle rheology} \label{sec:rheo}
Viscosity, $\eta$, is described as a function of mantle potential temperature, $T_m$, and water content, $c_w$:
\begin{equation}
	\eta = \left\{
	\begin{array}{ll}
		A \exp \kak{-\dfrac{E}{RT}} \kak{\Delta \eta_w}^{\kak{1 - c_w/c_0}} & (c_w \le c_0),\\
		A \exp \kak{-\dfrac{E}{RT}} \kak{\dfrac{c_0}{c_w}}& (c_w > c_0)
	\end{array}
	\right.
\end{equation}
where $E$ is an activation energy (300~kJ~mol$^{-1}$) and $R$ is the universal gas constant. Viscosity is assumed to decrease linearly with the water concentration $c_w$ \citep{Mei2000, Jain2019}, but an exponential relation is adopted below a cutoff value, $c_0 = 50$~ppm, to prevent viscosity from reaching infinity. CO$_2$ might also reduce mantle viscosity, but because it is mostly degassed before the solidification of the mantle surface (Figure~\ref{fig_degas}a), its effect is not considered. Viscosity contrast between dry and wet mantle of $c_w = c_0$ is assumed to be $\Delta \eta_w = 125$ \citep{Hirth1996, Mei2000}, and we adjust the preexponential constant, $A$, so that viscosity is $10^{19}$~Pa~s at 1350~$^\circ$C and $c_w = 0.04$~wt\%. Initial volatile concentrations in the mantle are chosen from Figures~\ref{fig_ocean}c and \ref{fig_Venus}b so that water oceans are absent at the beginning of solid-state convection stage.


\subsubsection{Mantle processing and crust formation} \label{sec:proc}
The processing of the mantle is assumed to start at a depth where the temperature exceeds a dry solidus. A wet mantle has a lower solidus temperature \citep[e.g.,][]{Kawamoto1997}, but the mantle would not be completely dehydrated until the melt fraction becomes sufficiently large, which becomes possible after crossing the dry solidus \citep[e.g.,][]{Hirth1996}. The initial depth of mantle processing can thus be approximated by the dry solidus, and we parameterize the initial pressure of melting, $P_i$ [GPa], as 
\begin{equation} \label{PiTm}
	P_i  = \frac{T_m - 1423~\mathrm{K}}{100},
\end{equation}
where the time-dependent term describes a decrease in the solidus temperature as the mantle becomes homogenized. The mixing timescale could be longer than 500~Myr, but it would not have a significant effect on the long-term evolution of surface water. As discussed in Section~\ref{sec:perc}, the mantle after magma ocean solidification likely has a heterogeneous structure including mostly consisted of high-Mg\# materials \citep{Elkins-tanton2008}, embedding Fe-rich blobs \citep{Miyazaki2019b}. An initially high-Mg\# mantle has a higher solidus than a pyrolitic mantle \citep{Miyazaki2022a}, but as the two components become homogenized by mantle mixing, the solidus temperature would decrease and approach that of the present day. 
The melting is assumed to stop where it reaches the base of the thermal boundary layer at $P_f = \rho_L g \kak{h_\mathrm{TBL} + h_c}$, where $\rho_L$ is the lithosphere density and $h_c$ is crustal thickness. The thickness of melting zone, $h_p$, can be described as
\begin{equation}
	h_p = \frac{P_i - P_f}{\rho_L g},
\end{equation}
and assuming that downwelling is much more localized than upwelling, the volumetric rate of mantle processing is given by
\begin{equation} \label{eq:Vproc}
 	\dot{V}_\mathrm{proc} = \frac{2 h_p u_\mathrm{conv}}{h_m} 4 \pi r_p^2.
\end{equation}
We adopt the scaling of \cite{Solomatov2000} for the average convective velocity beneath the stagnant lid:
\begin{equation}
	u_\mathrm{conv} = 0.38 \frac{\kappa}{h_m} \kak{\frac{Ra}{\theta}}^{1/2},
\end{equation}
where $\kappa$ is the thermal diffusivity, $Ra$ is the internal Rayleigh number, and $\theta$ is the Frank-Kamenetskii parameter. Finally, we can calculate the volumetric melt productivity by multiplying $\dot{V}_\mathrm{proc}$ by the average melt fraction in the partial melt layer, $\overline{\phi}$, which is estimated based on the melt productivity by adiabatic decompression ($d\phi/dP$=0.1~GPa$^{-1}$). The growth of crust is equivalent to the volumetric melt productivity, and the volume of DML increases by 
\begin{equation}
	\dot{V}_\mathrm{DML} = \kak{1-\overline{\phi}} \dot{V}_\mathrm{proc}.
\end{equation}
Volatiles included in the processed mantle are released to the mantle, and the rate of degassing is proportional to $\dot{V}_\mathrm{proc}$ as well. Volatiles are highly incompatible at the depths where melting occurs, so we consider that the processed mantle becomes entirely dry after melting. It is noted that, in previous studies, some large fraction of volatiles (90~\% of volatiles with 10~\% melting) was assumed to remain in the depleted mantle after melting \citep{Noack2017, Dorn2018}, which could underestimate the efficiency of degassing a factor of $\sim$10.  

\begin{figure}
\centering
	\begin{minipage}[l]{0.6\textwidth}
	\includegraphics[width=0.98\textwidth]{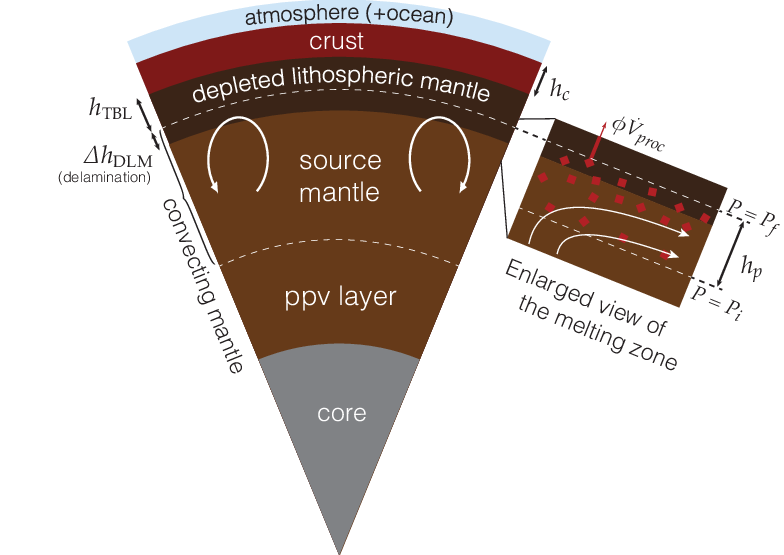}
	\end{minipage}
	\begin{minipage}[c]{0.37\textwidth}
	\caption{Schematic illustration of the structure of terrestrial planets, and parameters described in Section~\ref{sec:met3} are labeled. The initial basal pressure of depleted lithospheric mantle is set to $\sim$2~GPa, considering the temperature and composition of the mantle after magma ocean solidification. When the depleted lithospheric mantle layer is thicker than the thermal boundary layer, the excess layer would delaminate and be mixed with the source mantle. }
	\label{fig_methods}
	\end{minipage}
\end{figure}

Crustal thickness continues to grow by mantle processing, but the thickness of the depleted lithospheric mantle is reduced by delamination or rehydration from the underlying mantle. When the thermal boundary layer is thinner than DML ($h_\mathrm{DML} > h_\mathrm{TBL}$), DML would be eroded by convection (Figure~\ref{fig_methods}), so we assume that such a delaminated fraction of DML becomes mixed with the source mantle. Volatile concentrations in the mantle would be diluted as a result of mixing, and we adjust the concentrations accordingly \citep{Fraeman2010}. Also, DML is continuously rewetted by hydrogen diffusion from the underlying mantle. At each time step, a diffusion length $\Delta h_\mathrm{DML} = \sqrt{D_\mathrm{diff} \Delta t}$ is calculated, and we assume that the lower $\Delta h_\mathrm{DML}$ of DML is reincorporated into the source mantle reservoir. The diffusion coefficient is taken from a parameterization given in \cite{Korenaga2009}:
\begin{equation}
	D_\mathrm{diff} = 6 \times 10^{-5} \mathrm{m^2~s^{-1}} \times \kak{-0.0027 + 2.19 \times 10^{-6} \kak{T_m - 273}}.
\end{equation}
It is noted that diffusive rewetting is not considered when DML thickness is reduced by delamination. 

The initial thicknesses of crust $h_c$ and depleted lithospheric mantle $h_\mathrm{DML}$ are inherited from the last stage of magma ocean solidification. As discussed in Section~\ref{sec:perc}, the uppermost partially molten layer becomes depleted in volatiles through melt escape by percolation when the surface of the mantle solidifies (Figure~\ref{fig_cartoon}d). The bottom pressure of the initially processed mantle is set to $\sim$2~GPa, which is equivalent to 65~km for Earth-size planets and 90~km for 5$M_E$ super-Earths.

\subsubsection{Thermal evolution of the mantle} 
 The thermal evolution of the convecting mantle is controlled by a balance of radiogenic heating, heating from the core, convective heat flux of the mantle, and latent heat of mantle melting:
\begin{equation}
 	\rho_m C_m V_{cm} \gamma_m \frac{dT_m}{dt} = \dfrac{4}{3} \pi  \kak{r_m^3 - r_c^3} Q_m + 4\pi \kak{r_c^2 F_c - r_m^2 F_m} - \rho_L f_m L_m,
\end{equation}
where $\rho_m$ is the average density of mantle, $C_m$ is the specific heat of mantle, $V_{cm}$ is the volume of convective mantle, $r_m$ and $r_c$ are the radii of the mantle and core, $Q_m$ is radiogenic heat production per unit volume, and $L_m$ is the latent heat of mantle melting per unit mass. The constant $\gamma_m$ is adopted to convert potential temperature to the average temperature of the mantle. The convective heat flux of the mantle, $F_m$, can be calculated from $Nu$ (Section~\ref{sec:Nu}), and heat flux from the core, $F_c$, is derived using the scaling of \cite{Stevenson1983a}. 

The convecting mantle represents the whole mantle for planets smaller than Earth-size, but the deeper region of super-Earths may be too viscous for convection because its major constituting mineral, post-perovskite, is expected to have viscosity higher by a few orders of magnitude than the terrestrial mantle \citep{Tackley2013}. We thus assume that only the region shallower than $P_{ppv}$=200~GPa participates in thermal convection, and the value of $V_{cm}$ is adjusted accordingly. Also, the cooling of the deeper mantle would be inefficient, and heat flux supplied to the convecting mantle would be limited to those derived from core cooling and radiogenic decay \citep{Tackley2013, Dorn2018}, although core cooling would be much less efficient than the terrestrial mantle. With a viscosity of $10^{24}$~Pa~s assumed for the deep mantle, which is $\sim$$10^5$ higher than the terrestrial mantle, a local stability analysis of the bottom boundary layer \citep{Stevenson1983a} suggests that core cooling would be suppressed by $\sim$50 times for super-Earths.


\subsection{Results}
\subsubsection{Evolution of Earth-size planets} \label{sec:ME}
\begin{figure}
\centering
\includegraphics[width=\textwidth]{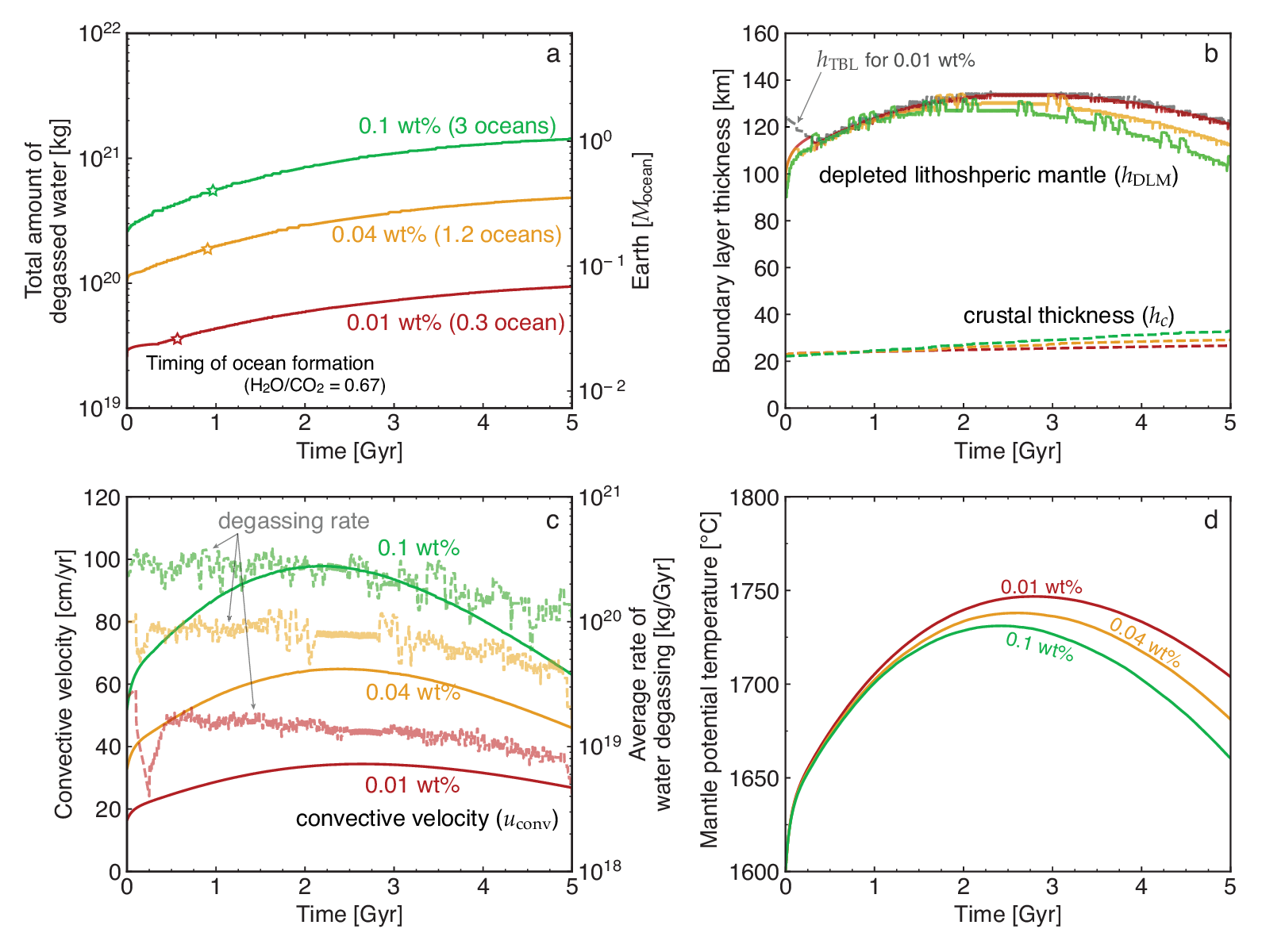}
\caption{The thermal evolution and degassing history of Earth-size planets under stagnant lid convection. (a) Total amount of degassed water, (b) thicknesses of crust (dashed), thermal boundary layer (solid) and depleted lithospheric mantle (gray), (c) convective velocity beneath the lid (solid) and the rate of water degassing averaged over 100~Myr (dashed, light-colored), and (d) mantle potential temperature are shown for three different initial H$_2$O concentrations: 0.01 (red), 0.04 (yellow), and 0.1~wt\% (green), corresponding to a total H$_2$O mass of 0.3, 1.2, and 3~ocean masses, respectively. Markers in (a) show the time when a threshold amount of surface H$_2$O is degassed to stabilize water oceans. The H$_2$O/CO$_2$ ratio of 0.66 (star) and net stellar radiation of 330~W~m$^{-2}$ are adopted here. With an H$_2$O concentration of 0.04~wt\%, $p_\mathrm{CO_2}$ corresponds to $\sim$330~bar. Once water oceans form at the surface, the mode of convection likely changes to plate tectonics, so modeling results after ocean formation would underestimate convective heat flux and the rate of mantle degassing. The gray line in (b) describes the thickness of thermal boundary layer. As a result of convective delamination, $h_\mathrm{DLM}$ becomes almost identical to $h_\mathrm{TBL}$ throughout the planetary evolution. }
\label{fig_stag1}
\end{figure}

Figure~\ref{fig_stag1} shows the thermal evolution and degassing history of a typical Earth-size planet, which illustrates that mantle degassing continues throughout its evolution (Figure~\ref{fig_stag1}a). The rate of mantle degassing can be understood from change in the thickness of the depleted lithospheric mantle $h_\mathrm{DML}$ because mantle processing is reflected in the growth of DML. During the first $\sim$1~Gyr, the depth where mantle melting starts becomes deeper with time (Figure~\ref{fig_stag1}b) as mantle temperature increases by radioactive elements, and also as the solidus temperature decreases as mantle mixing homogenizes heterogeneity created during magma ocean solidification (Equation~\ref{PiTm}). Therefore, the net growth of DML is observed, resulting in rapid degassing during this period (Figure~\ref{fig_stag1}c). In the subsequent stage, however, mantle temperature decreases with time, and the mantle is newly processed only as DML delaminates or is rewetted by hydrogen diffusion. The growth of DML by mantle melting and its delamination by convective erosion are occurring concurrently, and the two competing processes maintain $h_\mathrm{DML}$ comparable to $h_\mathrm{TBL}$ (Figure~\ref{fig_stag1}a). As a result, degassing is less efficient than in the first 1~Gyr, and its rate further slows down as the mantle cools down and convective velocity decreases (Figure~\ref{fig_stag1}c, \ref{fig_stag1}d). 

We ran the model for different initial water concentrations, and the results show that, under the same H$_2$O/CO$_2$ ratio, the timing of water ocean formation is nearly independent of the H$_2$O content ($x_\mathrm{H_2O}$; Figure~\ref{fig_stag1}a). This is a result of two competing processes. Because of a lower viscosity, convective velocity and thus the rate of mantle processing are higher for a wetter mantle: A mantle with a $\times$10 higher H$_2$O concentration has a $>$10~times higher H$_2$O degassing rate. 
Yet, the amount of surface H$_2$O necessary to stabilize oceans also becomes increasingly larger: An order of magnitude larger CO$_2$ concentration would require a surface H$_2$O amount larger by 40 times (Figure~\ref{fig_ocean}a). These two effects nearly cancel out, and thus an oceanless world persists for a similar duration as long as the H$_2$O/CO$_2$ ratio remains the same. Plate tectonics would operate once the threshold is reached \citep{Korenaga2020}, and mantle degassing would become more efficient afterwards. A wet surface is thus expected to be maintained during the subsequent evolution.

For an Earth-size planet, the atmosphere contains $\sim$5\% of the total H$_2$O budget immediately after the solidification of the mantle surface (Figures~\ref{fig_degas}a), and an additional $\sim$6\% would be degassed in the next 1~Gyr during solid-state convection (Figure~\ref{fig_stag1}a). The lower and upper bounds of these estimates, respectively, correspond to initial H$_2$O concentrations of 0.01 and 0.1~wt\%. For an H$_2$O/CO$_2$ mass ratio of 0.67, $\sim$10\% of the total H$_2$O budget needs to reside at the surface to stabilize water oceans, and such a condition is satisfied only after 0.5--1~Gyr of evolution (Figure~\ref{fig_stag1}a). Because degassing slows down with time, a planet with a smaller H/C ratio requires a longer duration to form water oceans. The H/C ratio therefore largely characterizes when oceans emerge on terrestrial planets.  

\begin{figure}
\centering
\includegraphics[width=\textwidth]{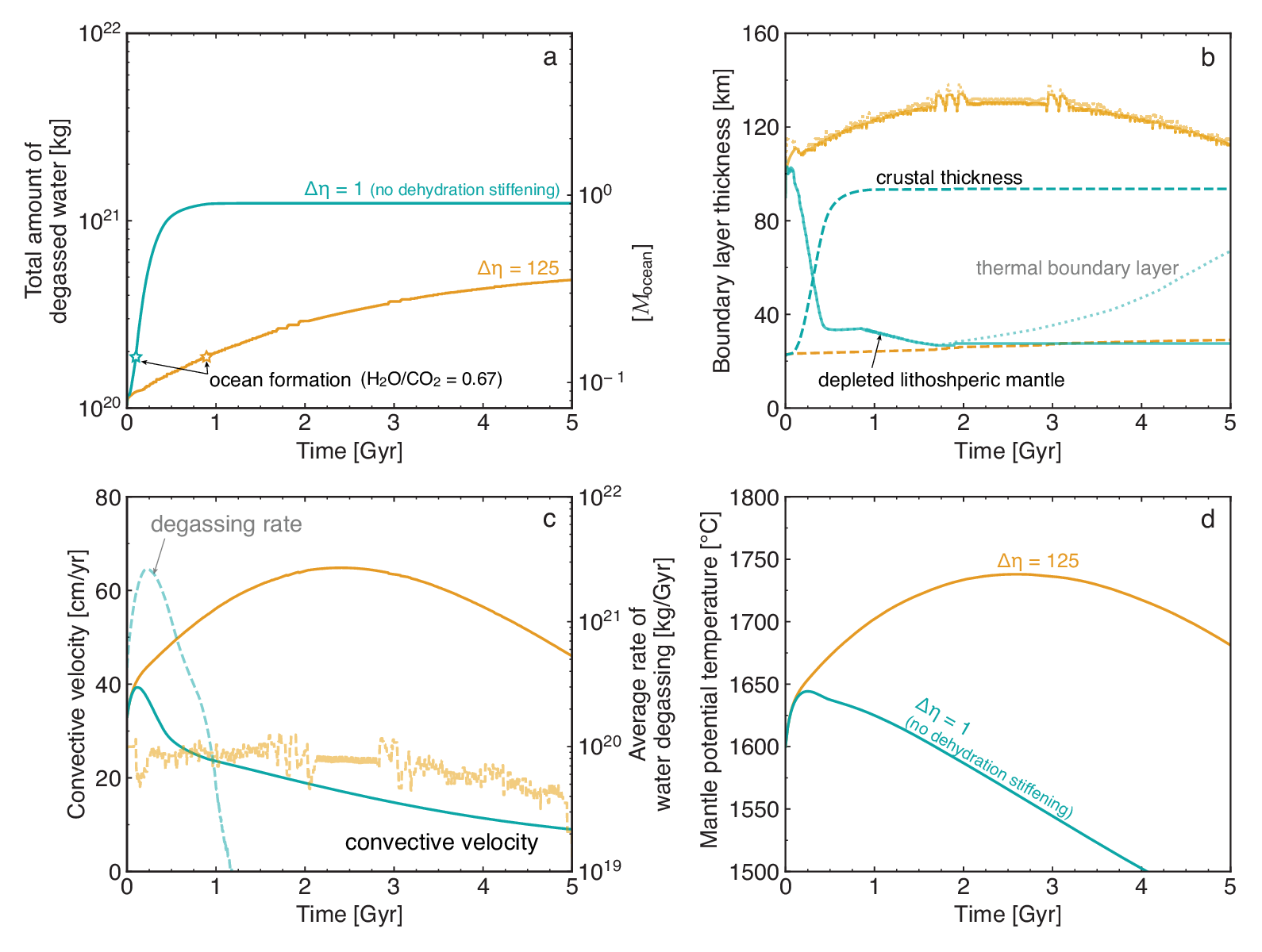}
\caption{The same as Figure~\ref{fig_stag1} but for an initial H$_2$O concentration of 0.1~wt\% with three different values for viscosity contrast between dry and wet mantle: $\Delta \eta_w=$ 1 (cyan) and 125 (yellow). A larger viscosity contrast results in slower degassing, and thus an oceanless surface would be maintained for a longer time. The case with $\Delta \eta_w=1$ does not exhibit any dehydration stiffening, and thus convection and mantle degassing are more efficient than the other two cases, although it is an unrealistic assumption. While depleted lithospheric mantle is actively eroded by thermal convection and its thickness is kept thin, crust grows rapidly to the point that, after 700~Myr, it becomes thick enough to halt mantle processing (shown in (c)). Markers in (a) show the time when a threshold amount of surface H$_2$O is degassed to stabilize water oceans.}
\label{fig_dehy}
\end{figure}

We note that a thick depleted lithospheric mantle acts as a limiting factor for heat transport and thus mantle degassing (Figure~\ref{fig_dehy}). In a conventional scaling of stagnant lid convection, which does not consider dehydration stiffening, a higher mantle temperature leads to a thinner thermal boundary layer and thus a higher convective heat flux to promote cooling \citep[the case of $\Delta \eta = 1$;][]{Solomatov2000}. Assuming such a scaling, the mantle is predicted to be efficiently processed from the beginning, crustal thickness would rapidly grow (Figure~\ref{fig_dehy}b), and water oceans would form within the first 100~Myr (Figure~\ref{fig_dehy}a). Neglecting dehydration stiffening thus severely overestimates the rate of mantle degassing. In reality, the presence of a dry DML impedes convection, and the thermal boundary layer becomes thicker at higher mantle potential temperatures (Figure~\ref{fig_stag1}a, \ref{fig_stag1}c). The Nusselt number $Nu$ is inversely proportional to $h_\mathrm{TBL}$, so a hotter mantle actually results in lower convective heat flux \citep{Korenaga2009}. Because of inefficient cooling, excess radiogenic heat production heats up the mantle during the first $\sim$1.5~Gyr (Figure~\ref{fig_stag1}c), and mantle temperature starts to decrease only after this period. The depth where mantle melting starts then becomes shallower with time, DML thins, and with a weaker influence of dehydration stiffening, $h_\mathrm{TBL}$ starts to decrease (Figure~\ref{fig_stag1}a). A larger viscosity contrast between dry and wet mantle $\Delta \eta_w$ further delays the rate of cooling and degassing (Figure~\ref{fig_dehy}c), and thus better constraining the dependence of viscosity on water content would be important to accurately predict the timing of ocean formation. Some experimental studies suggest that the water dependence of mantle viscosity can be more drastic than assumed here \citep{Faul2007}.

\subsubsection{Evolution of super-Earths} \label{sec:sup}
\begin{figure}
\centering
\includegraphics[width=\textwidth]{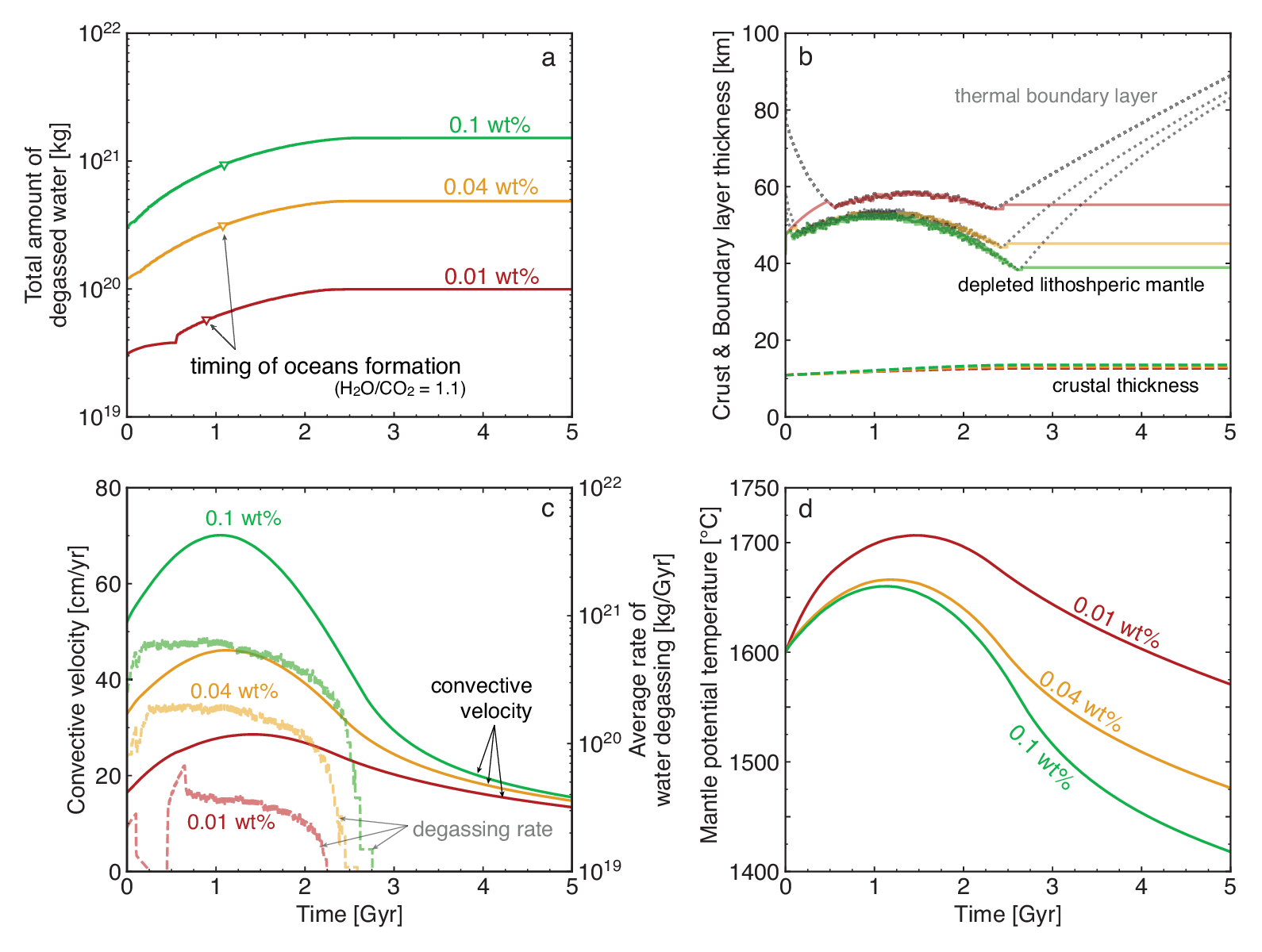}
\caption{The thermal evolution and degassing history of  5$M_E$ super-Earths under stagnant lid convection. (a) Total amount of degassed water, (b) thicknesses of crust (dashed), thermal boundary layer (solid) and depleted lithospheric mantle (gray), (c) convective velocity beneath the lid (solid) and the rate of water degassing averaged over 100~Myr (dashed, light-colored), and (d) mantle potential temperature are shown for four different initial H$_2$O concentrations: 0.01 (red), 0.04 (yellow), 0.1 (green), and 0.4~wt\% (cyan). Markers in (a) show the time when a threshold amount of surface H$_2$O is degassed to stabilize water oceans, where the H$_2$O/CO$_2$ ratios of 1.1  and net stellar radiation of 330~W~m$^{-2}$ are assumed. The ratios adopted here are higher than those in Figures~\ref{fig_stag1} and \ref{fig_dehy}, and with H/C=0.45, oceans are absent throughout the evolution for H$_2$O concentrations higher than 0.04~wt\%. For a H$_2$O concentration of 0.04~wt\%, H$_2$O/CO$_2$$=1.1$ corresponds to $p_\mathrm{CO_2} \sim$500~bar, and the total H$_2$O inventory is 6~ocean masses because the mantle mass of a 5$M_E$ super-Earth is $\sim$5 times that of Earth.}
\label{fig_stag2}
\end{figure}

\begin{figure}
\centering
	\begin{minipage}[l]{0.6\textwidth}
	\includegraphics[width=.96\textwidth]{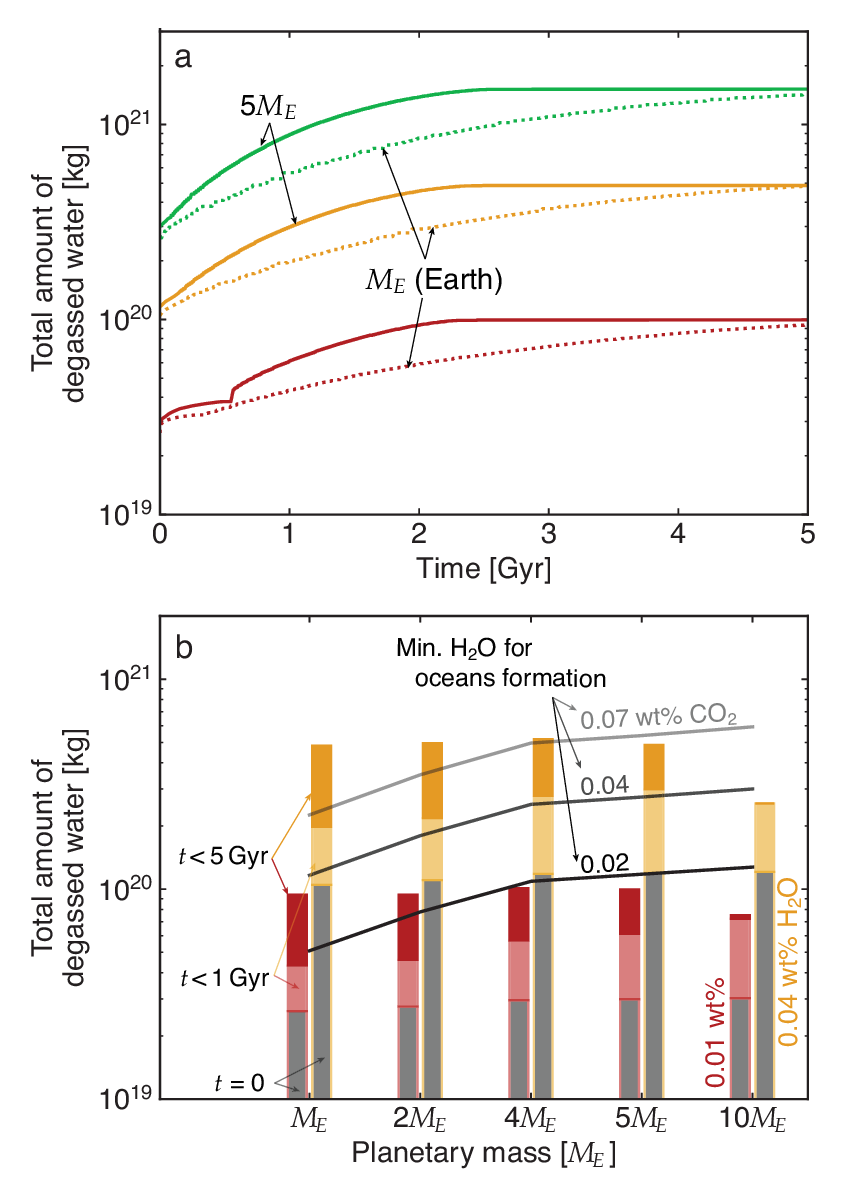}
	\end{minipage}
	\begin{minipage}[c]{0.38\textwidth}
	\caption{(a) The total amount of degassed water for Earth-size (solid) and 5$M_E$ super-Earth planets (dotted) with three different initial H$_2$O concentrations in the mantle $x_\mathrm{H_2O}$: 0.01 (red), 0.04 (yellow), and 0.1~wt\% (green). Mantle degassing is faster during the first 1~Gyr but ceases after $\sim$2~Gyr for a 5$M_E$ super-Earth. (b) The amount of degassed water for different planetary sizes with H$_2$O concentrations of 0.01 (red) and 0.04~wt\% (yellow). The bars show the amount of degassed water during three different periods: before the solidification of a magma ocean (gray), before 1~Gyr (light red/yellow) and 5~Gyr of solid-state convection (dark red/yellow). The minimum amount of surface H$_2$O for ocean formation is also plotted for initial CO$_2$ concentrations $x_\mathrm{CO_2}$ of 0.02 (black), 0.04 (dark gray), and 0.07~wt\% (light gray).}
	\label{fig_ratio}
	\end{minipage}
\end{figure}

For larger planets, the formation of water oceans is predicted only under a restricted range of initial conditions, and even when oceans are expected to emerge, the formation would take longer than Earth-size planets (Figure~\ref{fig_stag2}a). As discussed in Section~\ref{sec:res2}, more CO$_2$ is released during the magma ocean stage on larger planets (Figure~\ref{fig_ocean}b), so a greater amount of surface H$_2$O is required to stabilize oceans (Figure~\ref{fig_ocean}a). Yet, the amount of water residing at the surface is nearly independent of planetary mass at the beginning of the evolution (Figure~\ref{fig_ocean}a), and the amount of degassed volatiles during solid-state convection increases little with planetary size (Figure~\ref{fig_ratio}). Under the same initial volatile concentrations, larger planets thus may not have a sufficient amount of surface H$_2$O to stabilize oceans.

Larger planets have a higher rate of degassing, but mantle processing ceases earlier during solid-state convection. This is seen in Figure~\ref{fig_ratio}a, which compares the evolution of surface water for Earth-size and 5$M_E$ super-Earth planets: the total amount of degassed water increases faster during the first 1--2~Gyr for a 5$M_E$ super-Earth, yet no degassing is observed during the subsequent period. The scaling for mantle processing can be derived from Equation~(\ref{eq:Vproc}):
\begin{equation} \label{eq:scale}
	V_\mathrm{proc} \sim f_m \Delta t \sim \frac{h_p u_\mathrm{conv} r_p^2}{h_{cm}} \Delta t,
\end{equation}
where $h_{cm}$ is the depth of convecting mantle. The scaling for whole mantle convection ($h_{cm}$=$h_m$) has been provided in \cite{Orourke2012} as $V_\mathrm{proc} \sim M_p^{0.24} \Delta t$, and a similar scaling for super-Earths can be derived considering the effect of viscous post-perovskite in the deeper mantle: $h_{cm}$ decreases to $P_{ppv}/\kak{\rho_m g}$, and $u_\mathrm{conv}$ to $Ra^{1/2}/h_{cm} \sim \kak{h_{cm} g}^{1/2} \sim 1$. Using the interior model of \cite{Valencia2006} $\kak{r_p \sim M_p^{0.26}}$, $V_\mathrm{proc}$ follows $r_p^2 \Delta t \sim M_p^{0.52} \Delta t$, and degassing is in general more efficient for larger planets during the first 1~Gyr of continuous degassing (Figure~\ref{fig_ratio}). The duration of degassing, $\Delta t$, however, shortens with increasing planetary size. Mantle processing is triggered when the initial depth of melting, $h_i$=$P_i/\kak{\rho_L g}$, is greater than the base of the thermal boundary layer $h_\mathrm{TBL}$, but larger planets are less likely to meet such a condition. When super-Earths of different sizes under the same thermal state are considered, the initial depth of melting scales with $g^{-1}$, whereas the thickness of the thermal boundary layer is scaled as
\begin{equation}
 	h_\mathrm{TBL}  = \frac{h_{cm}}{Nu} \sim \frac{h_{cm}}{Ra^{1/3}} \sim g^{-1/3}.
\end{equation}
With increasing planetary mass, the initial melting depth becomes shallower at a faster rate than TBL, and triggering mantle processing becomes more difficult. 
Consequently, mantle processing ceases earlier for larger planets, and even though the rate of mantle degassing is higher during the early stage, the total amount of degassed volatiles throughout their evolution does not change appreciably with planetary size (Figure~\ref{fig_ratio}b). We note that the limited processing of the mantle for larger planets is in a broad agreement with previous studies of 2-D mantle convection models \citep{Dorn2018}. A larger fraction of the water budget is thus expected to remain in the mantle for super-Earths without degassing to the surface, precluding the formation of water oceans.

\subsection{Discussion: Hydrogen escape and long-term habitability} \label{sec:dis3}
\begin{figure}
\centering
	\begin{minipage}[l]{0.6\textwidth}
	\includegraphics[width=.96\textwidth]{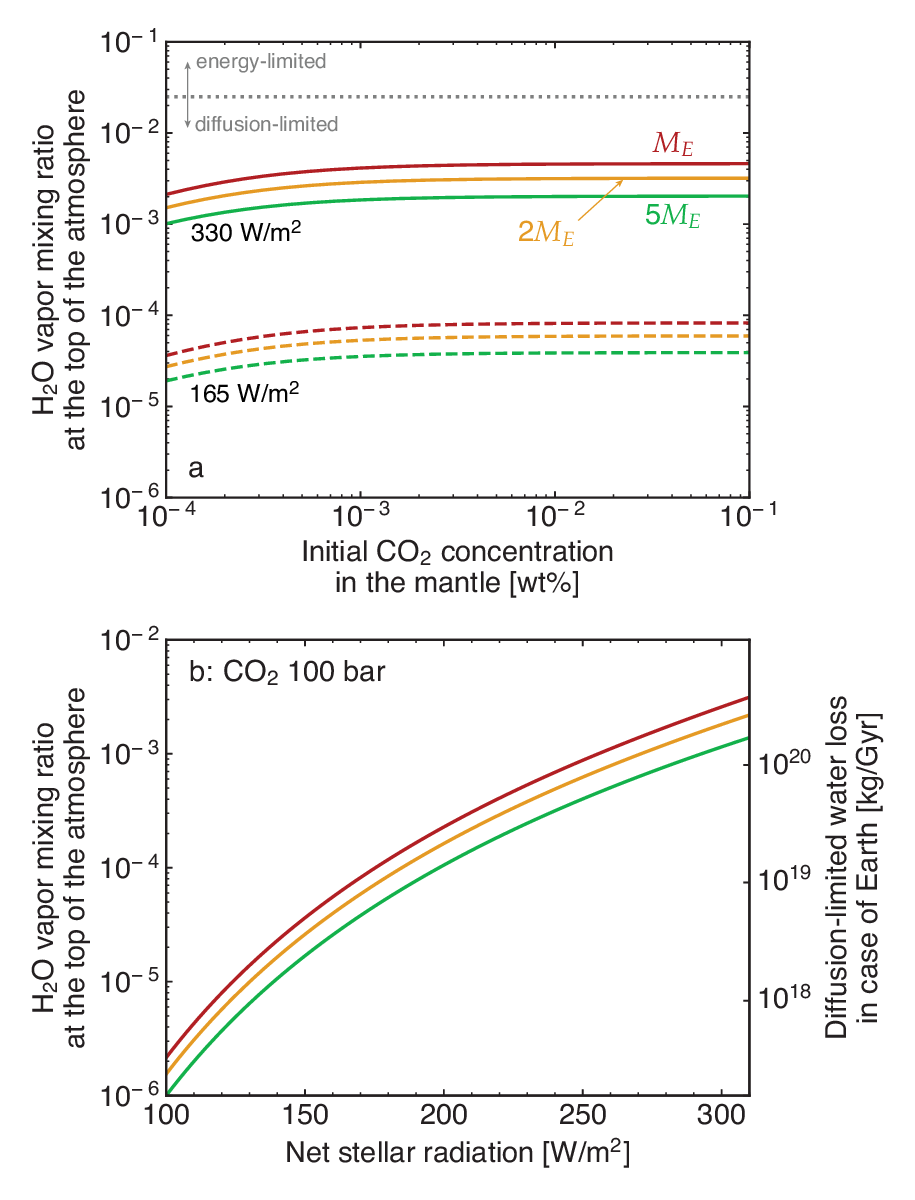}
	\end{minipage}
	\begin{minipage}[c]{0.38\textwidth}
	\caption{The mixing ratio of H$_2$O vapor at the top of the atmosphere as a function of (a) the initial CO$_2$ concentration in the mantle and (b) as of net stellar radiation. A planet with water oceans is assumed, and an oceanless one would have a smaller mixing ratio than shown here. Colors indicate planetary mass: Earth-size (red), 2$M_E$ (yellow), and 5$M_E$ super-Earth (green). (a) Solid and dashed lines represent net stellar radiation of 330 and 165~W~m$^{-2}$, respectively. Dotted gray line describes a critical ratio above which hydrogen escape is limited by the magnitude of EUV flux, and for the range of stellar radiation considered here, hydrogen escape is limited by diffusion. (b) An atmosphere containing 1~bar of N$_2$ and 100~bar of CO$_2$ is assumed here. Water loss by diffusion-limited hydrogen escape is estimated as well for an Earth-size planet using a binary diffusion coefficient of $b_\mathrm{H_2}$=$1.46 \times 10^{21}$~m$^{-1}$~s$^{-1}$ \citep{Catling2017}. Water is assumed to readily convert to hydrogen at the top of the atmosphere, and thus the values should be treated as an upper bound estimate for water loss. }
	\label{fig_mixing}
	\end{minipage}
\end{figure}

Hydrogen escape can be important for the timing of ocean formation when the initial H$_2$O concentration in the mantle $\kak{x_\mathrm{H_2O}}$ is small and the planet is close to the star. The efficiency of hydrogen escape depends on the mixing ratio of water vapor at the top of the atmosphere, and diffusion is the rate-controlling process when the mixing ratio is below $\sim$2--3\%, assuming a similar EUV flux to the present-day Sun and a heating efficiency of 30\% \citep{Watson1981}. Although the luminosity of younger stars may be higher than the present-day Sun, hydrogen escape from a planet without a significant envelope would always be limited by diffusion, and thus a larger flux size would not affect the rate of water loss. The 1-D radiative-convective model \citep[Appendix~\ref{sec:atmos}; ][]{Nakajima1992} predicts that the mixing ratio is smaller than 1\% for conditions considered in this study (Figure~\ref{fig_mixing}), and thus hydrogen escape would be limited by diffusion \citep[e.g.,][]{Catling2017}. For a 5$M_E$ super-Earth with a water vapor mixing ratio of 0.003, the total amount of hydrogen escape during a 1~Gyr period is estimated to be 4.3$\times 10^{19}$~kg, which is equivalent to 3.8$\times 10^{20}$~kg of H$_2$O. This amount is comparable to the total amount of degassed water for $x_\mathrm{H_2O}$$<$0.04~wt\% (Figure~\ref{fig_ratio}a), and thus ocean formation would be delayed, or in some case precluded, compared to what is predicted in Figures~\ref{fig_stag1}a and \ref{fig_stag2}a. Diffusion-limited flux is linearly proportional to the mixing ratio, so under a weaker net stellar radiation, the amount of water lost by hydrogen escape is negligible because the atmosphere contains a smaller amount of water vapor (Figure~\ref{fig_mixing}b).

\begin{figure}
\centering
\includegraphics[width=1.01\textwidth]{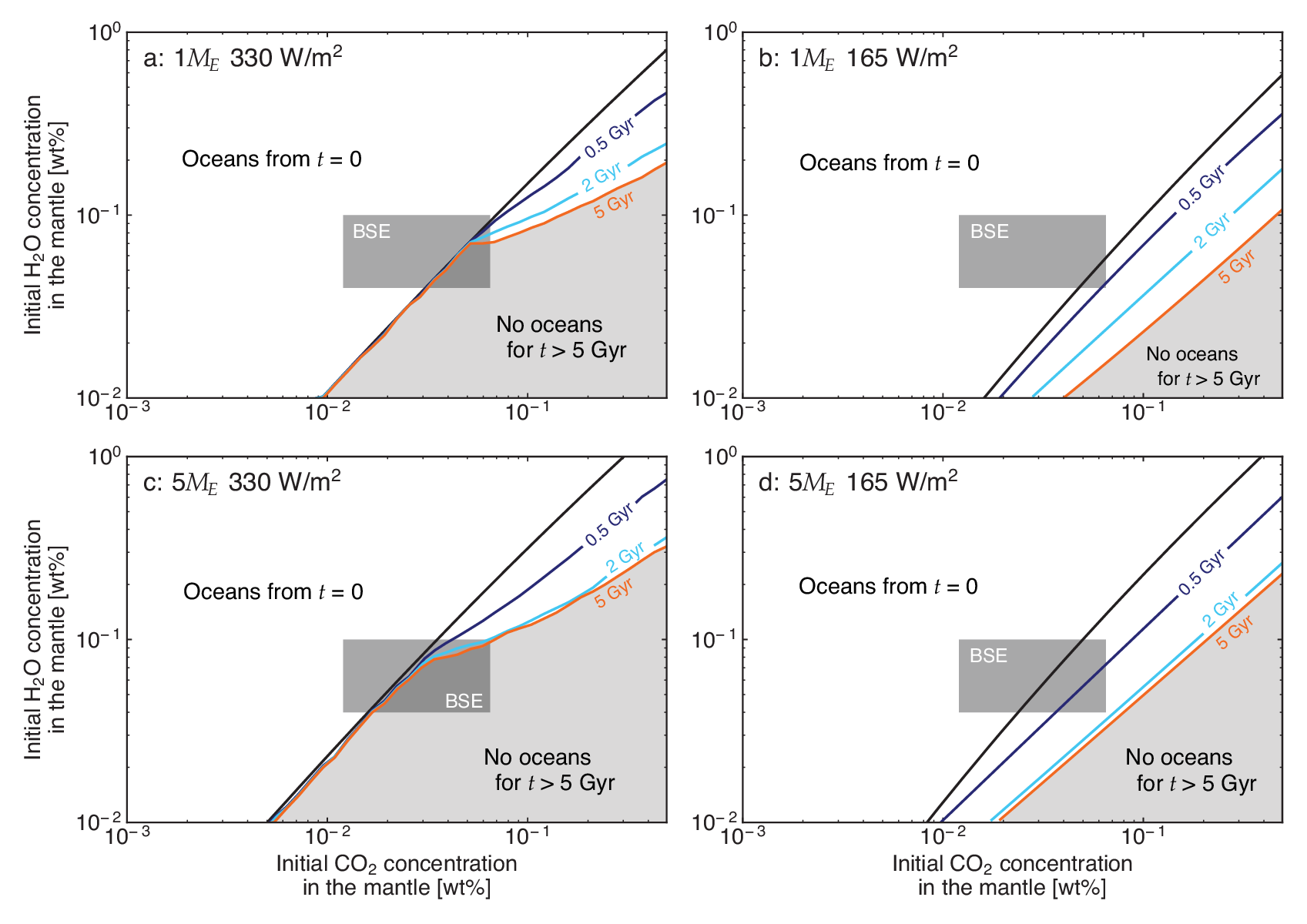}
\caption{Contours showing the timing of ocean formation for Earth-size (top) and 5$M_E$ super-Earth planets (bottom). Results are calculated for two different levels of net stellar radiation: 330 (left) and 165~W~m$^{-2}$ (right). Water oceans form immediately after magma ocean solidification and exist throughout the evolution of planets in regions left of black lines, whereas oceans would be absent for longer than 5~Gyr under conditions shaded in light gray. Rectangles represent the estimated volatile concentrations in the bulk silicate Earth (BSE) from \cite{Hirschmann2009} and \cite{Korenaga2017b}. }
\label{fig_contour}
\end{figure}

We calculate mantle degassing under solid-state convection and estimate the likelihood and timing of ocean formation for various initial volatile concentrations, considering the loss of water by hydrogen escape. With a stronger net stellar radiation and thus a higher water mixing ratio, hydrogen escape becomes significant and prohibits the formation of ocean formation at a small initial H$_2$O concentration $\kak{x_\mathrm{H_2O}}$. On early Venus, for example, if oceans are absent at the time of magma ocean solidification, degassing during solid-state convection does not supply a sufficient amount of water to stabilize oceans, and the surface would lack oceans throughout its evolution under $x_\mathrm{H_2O}$$<$0.08~wt\%, equivalent to a total water inventory of $\sim$2.4~ocean mass (Figure~\ref{fig_contour}a). Although the threshold $x_\mathrm{H_2O}$ is higher for larger planets (Figure~\ref{fig_contour}d), the general trend remains the same. On the other hand, if $x_\mathrm{CO_2}$ is small and thus water oceans are formed from the beginning, the mantle would operate under plate tectonics, and efficient mantle degassing would maintain oceans during the subsequent period. Therefore, for planets receiving $>$240~W~m$^{-2}$ of radiation with a small initial H$_2$O content, whether water oceans exist immediately after magma ocean solidification characterizes the long-term existence of water oceans and thus the habitability of planets (Figure~\ref{fig_contour}). 

This could provide an explanation for why Venus and Earth had divergent evolutionary paths: Venus may have continued to lack oceans throughout its history because of the comparable rates of mantle degassing and hydrogen escape, whereas Earth has had water oceans and plate tectonics since the solidification of the mantle surface. The net solar radiation received by early Venus is considered to be close to the tropospheric radiation limit \citep{Hamano2013}, but even if Venus was not in a runaway greenhouse state, water oceans are not guaranteed on early Venus. Limited mantle degassing under stagnant lid convection also implies that the Venusian mantle could still be wet, and that the amount of hydrogen escape and thus its by-product oxygen could be smaller than previously predicted. Such a small amount of oxygen could be consumed by the oxidation of surface rocks without invoking other removal mechanisms \citep[e.g.,][]{Kurosawa2015}. 

Hydrogen escape has little influence on ocean formation when the stellar radiation is weaker and/or a mantle contains a greater amount of H$_2$O (Figure~\ref{fig_contour}c, \ref{fig_contour}f). In such cases, planets with the same H/C ratio would develop water oceans on a similar timescale regardless of the absolute value of initial H$_2$O concentrations as discussed in Section~\ref{sec:ME}. For planets resembling early-Earth, oceans would exist from the beginning for H/C$>$0.69, and the formation timescale becomes increasingly longer as the H/C ratio decreases because degassing slows down as the mantle cools (Figures~\ref{fig_stag1}c, \ref{fig_stag2}c; Section~\ref{sec:sup}). Degassing eventually discontinues, and thus the surface would never acquire a sufficient amount of water to stabilize oceans for H/C ratios lower than 0.16. Such threshold H/C ratios for ocean formation increase with planetary mass and net stellar radiation. A 5$M_E$ super-Earth receiving the same radiation as early Earth would not develop oceans for H/C$<$0.37 (Figure~\ref{fig_contour}f), and its threshold increases to $\sim$0.7 if the planet is located at the orbit of early Venus (Figure~\ref{fig_contour}d). These values are within the estimate for the bulk silicate Earth \citep[0.99 $\pm$ 0.42;][]{Hirschmann2009}, so if terrestrial exoplanets have similar volatile concentrations to Earth, the presence of water oceans is not guaranteed, especially for super-Earths. Planets may have a different fate with a slight difference in their volatile concentrations, and the habitability of planets would depend largely on their size and the H/C ratio of the mantle, in addition to the distance from the central star.

Although its effect is not considered in our study, the rate of hydrogen escape during the magma ocean phase is much higher than discussed here. A hot surface allows most of the surface water budget to exist as vapor, increasing the mixing ratio at the top of the atmosphere. Because a molten surface only persists for less than 0.1 Myr in a period after the final giant impact, water loss by hydrogen escape is negligible in the scope of Section 2. Hydrogen escape during the accretionary phase, however, could lead to a significant water loss \citep{Zahnle1988} and would be important when constraining the water budget of terrestrial planets from the perspective of the chemical composition of planetary building blocks.

\section{Discussion and Conclusions}
\subsection{Additional processes to promote mantle degassing}

During the transition from a magma ocean to solid-state convection, we assumed that the solid-dominated layer quickly solidified by the Rayleigh-Taylor instability within the layer. As discussed in Section 2.1, the density structure of the solid-dominated layer is gravitationally unstable, so a newly-solidified cold material would continuously be mixed with the pre-existing solid-dominated layer \citep{Maurice2017}. Because the cooling of the surface melt-dominated layer is much faster than the solid-dominated one, the thermal evolution of the solid-dominated layer would mostly be governed by the cold downwelling of newly-solidified materials, which is likely to be more efficient than cooling by solid state convection \citep{LeBrun2013, Zahnle2015}. Its effect, however, remains unquantified because its modeling requires the implementation of solidification by adiabatic compression. Viscosity also changes drastically as partially molten materials solidify during the downwelling. Fully incorporating these effects would require some numerical efforts, so we cannot rule out the possibility that the solid-dominated layer remains superadiabatic at the time of surface solidification. In such a case, remaining heat may be released by episodic eruptions, and the amount of surface water could be slightly higher than predicted in Section 2.

Mantle degassing may also be promoted by tidal heating. It has been suggested that tidal heating inhibits the cooling of a solid-dominated magma ocean \citep{Zahnle2015}, and if the cooling timescale becomes long enough, the percolation of pore melt may transport a significant amount of trapped volatiles to the surface. Quantifying the timescale of mantle cooling, however, is not straightforward. As discussed in the previous paragraph, the cooling of the solid-dominated layer is mostly controlled by the overlying melt-dominated layer, and any heat added by tidal dissipation could efficiently be transported upwards by the Rayleigh-Taylor instability within the layer. A simple 1-D model would be insufficient to address this complex transition period, and further investigation is warranted. Also, the presence of satellites is not always guaranteed, and even if a satellite exists, its size is not necessarily going to be as large as the Earth's moon. The results shown here would thus serve as a baseline understanding on volatiles degassing, and we plan to address additional processes in future studies.

\subsection{Plate tectonics and the removal of thick CO$_2$ atmosphere} \label{sec:hab}
Plate tectonics maintains a habitable atmospheric composition over geological time \citep{Berner2004}, so the presence of surface water, a key to plate tectonics, is crucial for the long-term habitability of the planet. Its role is particularly important during the early stage of evolution because the majority of the CO$_2$ budget is degassed during magma ocean solidification. 
When $p_\mathrm{CO_2}$$>$100~bar, its greenhouse effect would maintain a surface temperature over 100~$^\circ$C \citep{Abe1988, Abe1993a}, which may prevent life forms similar to those existed on Earth from evolving. The removal of such a thick CO$_2$ atmosphere likely requires rapid plate motion \citep{Sleep2001a, Miyazaki2022a}, so although this study has focused on the formation of water oceans, stabilizing water oceans is only one of the requirement for creating a habitable environment. 

The efficiency of CO$_2$ removal from the atmosphere depends on plate velocity, and when the mantle has a pyrolitic composition, plate velocity could be slower for a hotter mantle than in the present-day \citep{Korenaga2006, Korenaga2010}. \cite{Miyazaki2022a} suggests that a chemically heterogeneous mantle resulting from magma oceans solidification would develop a thinner depleted lithospheric mantle and thus allows faster plate velocity than a pyrolitic mantle. Whether such chemical differentiation occurs in a magma ocean can be a critical factor in developing a habitable environment, although it depends on several unconstrained physical properties and is subjected to debate \citep{Solomatov1993a, Solomatov1993, Xie2020}. 


Plate velocity also depends strongly on the degree of mantle hydration \citep{Korenaga2010}, and the idea of a wet mantle after magma ocean solidification is also consistent with the removal of the thick CO$_2$ atmosphere. The previous models of magma ocean degassing have often assumed that the mantle would be entirely dehydrated during solidification \citep{Elkins-tanton2008, Hamano2013, LeBrun2013, Salvador2017}, but a dry mantle would have a slower plate motion than in the present-day, despite the hotter mantle temperature in the past \citep{Korenaga2006}. If so, the removal of CO$_2$ over $>$100~bar would become highly inefficient \citep{Miyazaki2022a}, which fails to explain the emergence of the moderate climate by the early Archean. 
The wet mantle and thus rapid plate motion also agree with recent geochemical modeling, which suggests the efficient recycling of the surface material in the Hadean \citep{Rosas2018, Hyung2020}. Studies on magma ocean have often suffered from a lack of observational constraints, but we may be able to obtain new insights into this period by taking into account the aftermath of a magma ocean.

\subsection{Conclusions}
We estimated the likelihood of ocean formation and its timing for planets with various volatile concentrations by solving for the evolution of mantle degassing during and after magma ocean solidification. Considering the rheological transition of a partially molten medium, magma ocean is predicted to retain a significant fraction of the H$_2$O inventory, whereas CO$_2$, which is less soluble to magma, would mostly reside in the atmosphere after the solidification. When the H/C ratio is lower than a threshold, surface water exists entirely as water vapor as a result of a strong greenhouse effect, and water oceans would be absent on the surface. Larger planets are less likely to develop water oceans because more greenhouse gases are emitted to the atmosphere during the solidification process.

If water oceans do not form at the beginning of solid-state convection, planets would operate under stagnant lid convection, severely limiting mantle degassing during the subsequent evolution. The rate of mantle degassing decreases and eventually reaches zero as the mantle cools and the thermal boundary layer grows, so the majority of H$_2$O is expected to be retained in the mantle throughout its evolution. The total amount of degassed water is nearly independent of planetary mass under the same volatile concentrations, but because a CO$_2$ atmosphere is thicker and thus a larger amount of surface water is necessary to form oceans for larger planets, the threshold H/C ratio for ocean formation increases with planetary size. The threshold also increases with net stellar radiation as a larger amount of water vapor is contained in the atmosphere. In addition, water loss by hydrogen escape may preclude ocean formation for planets receiving strong radiation and containing a small amount of H$_2$O. These would not be the case once water oceans form at the surface because the mode of mantle convection would likely change to plate tectonics, promoting mantle degassing during the subsequent evolution.

 
If the volatile contents of rocky exoplanets are similar to those of the bulk silicate Earth, mantle degassing during magma ocean would not supply a sufficient amount of H$_2$O when planets are larger and/or receive stronger radiation than Earth. Exoplanets could be richer in volatiles than Earth, and indeed, exoplanets of larger size are known to have an atmosphere of smaller metallicity \citep{Benneke2019a}. Yet, the redox evolution of rocky planets is not well understood, and if hydrogen existed in a reduced form, a high H/C ratio may not necessarily result in the formation of water oceans. An increasing number of observations for the exoplanet atmosphere is expected with upcoming missions, but the detection of water vapor does not necessarily indicate the presence of water oceans. In addition to the habitable zone, planetary mass and the H$_2$O/CO$_2$ ratio are also important parameters in predicting the likelihood of oceans on exoplanets. Because the threshold H/C ratio for ocean formation is lower for smaller planets receiving weaker radiation, such planets are a more promising candidate to search for potential life, although they may be more difficult to detect and observe the atmosphere. 
 




\acknowledgments
This work was sponsored by the U.S. National Aeronautics and Space Administration under Cooperative Agreement No. 80NSSC19M0069 issued through the Science Mission Directorate and the National Science Foundation under grant EAR-1753916, and also in part by the facilities and staff of the Yale University Faculty of Arts and Sciences High Performance Computing Center. Y.~M. was supported by the Stanback Postdoctoral Fellowship from Caltech Center for Comparative Planetary Evolution. This manuscript benefited from the comments by Dan Bower and Darius Modirrousta-Galian.

\appendix
\section{Method details}
\subsection{The thermal structure of the melt-dominated layer}
Mantle potential temperature, $T_m$, is calculated to be consistent with the depth of the melt-dominated layer $d$. The magma ocean is considered to be adiabatic as a result of rapid convection, and $T_m$ is obtained by integrating the adiabatic temperature gradient from the bottom of the melt-dominated layer to the surface. The adiabatic gradient is described as
\begin{equation}
 	\left( \frac{dT}{dP} \right)_S = \frac{\alpha' T}{\rho c_p'},
\end{equation}
where thermal expansivity, $\alpha'$, and specific heat, $c_p'$, include the effect of phase transition. They are obtained by differentiating the Gibbs free energy of peridotite described in \cite{Miyazaki2019b}, and this thermodynamic model is also used to determine other thermodynamic properties, such as density and melt fraction at each depth. The relation between $T_m$ and $d$ for an Earth-size planet is shown in Figure~2a.

The surface temperature $T_s$ is calculated using a 1-D atmospheric model (Appendix~\ref{sec:atmos}) and depends on the atmospheric mass and composition, calculated from Equations (1-3), and the net outgoing infrared radiation ($F_\uparrow (\tau=0)$ in Appendix A.2). The outgoing infrared radiation is the sum of absorbed stellar radiation and convective heat flux, $F_\mathrm{conv}$, given by \citep{Zahnle1988, Solomatov2007}:
\begin{equation}
 	F_\mathrm{conv} = 0.089 \frac{k (T_m - T_s)}{d} Ra^{1/3},
\end{equation}
where $k$ is thermal conductivity and $Ra$ is the Rayleigh number of the melt-dominated layer. Equation (A2) assumes that heat flux from the solid-dominated layer can be ignored because of a much higher viscosity than the melt-dominated one. The Rayleigh number is defined as 
\begin{equation}
	Ra= \frac{\alpha \rho g (T_m - T_s) d^3}{\kappa \eta_l},
\end{equation}
where $\alpha$ is thermal expansivity, $\kappa$ is thermal diffusivity, and $\eta_l$ is the viscosity of the melt-dominated layer. The viscosity $\eta_l$ does not change significantly until magma experiences rheological transition and thus is set to a constant of 1000 Pa s. The net stellar radiation would be $\sim$167~W~m$^{-2}$ for the early Earth and $\sim$319 W m$^{-2}$ for the early Venus with an albedo of 0.3 and a solar radiation 30\% weaker than the present day.

\subsection{Atmospheric model} \label{sec:atmos}
The surface temperature and the total amount water vapor in the atmosphere are estimated using a 1-D radiative-convective model by \cite{Nakajima1992}. We consider two layers within the atmosphere: stratosphere and troposphere, where the temperature is controlled by different heat transport mechanism. Whereas radiative equilibrium governs the stratosphere, the thermal structure in the troposphere is characterized by convective heat transport from the bottom. 
The atmosphere is assumed to be plane-parallel and is transparent to stellar radiation, but opaque to infrared radiation regardless of the wavelength. Although this assumption of gray atmosphere may underestimate the surface temperature during the earlier stage of magma ocean, the calculated temperatures are similar for the gray and nongray cases for surface temperatures below 200~$^\circ$C \citep{Salvador2017}. For simplicity, the effects of clouds are neglected in our model. 

The thermal profile of the stratosphere can be written as a function of optical thickness, $\tau$:
\begin{equation}
	\sigma_B T(\tau)^4 = \frac{1}{2} F_\mathrm{net} \left( \frac{3}{2} \tau + 1 \right),
\end{equation}
where $\sigma_B$ is the Stefan-Boltzmann constant and $F_\mathrm{net}$ is the net infrared flux emitted from the top of the atmosphere. Upward, $F_\uparrow$, and downward radiation fluxes, $F_\downarrow$, are also calculated as
\begin{eqnarray}
	F_\uparrow(\tau) &=& \frac{1}{2} F_\mathrm{net} \left( \frac{3}{2} \tau + 2 \right), \label{rad}\\
	F_\downarrow(\tau) &=& \frac{1}{2} F_\mathrm{net} \left( \frac{3}{2} \tau \right).
\end{eqnarray}
Optical thickness $\tau$ is defined so that it increases towards the Earth's surface:
\begin{equation}
	\tau(z) = - \int_{\infty}^{z} \kappa \rho_g dz.
\end{equation}
Opacity, $\kappa$, depends on the atmospheric concentrations of greenhouse gases, and for the atmosphere which has a molar fraction of $x_i$ of gas species $i$, $\kappa$ is given by
\begin{equation}
	\kappa = \frac{1}{\mu} \sum_i \kappa_i x_i \mu_i,
\end{equation}
where $\mu$ is the mean molar mass of gas, $\kappa_i$ and $\mu_i$ are the opacity and molar mass, respectively, of species $i$. For an infrared absorption coefficient, we use $10^{-2}$ for H$_2$O, $1.3 \times 10^{-4}$ for CO$_2$ \citep{Abe1985}, and 0 for N$_2$, assuming N$_2$ is transparent to infrared radiation. The composition of the atmosphere is assumed to be uniform in the stratosphere. 

In the troposphere, the thermal structure is controlled by the moist adiabatic lapse rate, which is described as
\begin{equation}
	\left( \frac{dT}{dP} \right) = \frac{\mu}{\rho_g C_p} \frac{1 + \dfrac{qL}{RT}}{1 + \dfrac{qL^2}{C_p RT^2}},
\end{equation}
where $\rho_g$ is the mean gas density, $C_p$ is heat capacity, $q$ is water mixing ratio, and $L$ is the latent heat of the water. Water mixing ratio $q$ is set to be saturated, which is likely in a CO$_2$ dominated atmosphere, and is obtained from the water phase diagram. 

The tropopause, a boundary between the stratosphere and troposphere, is determined as a height that the radiation energy is balanced. The upward radiation flux emitted from the troposphere can be calculated as
\begin{equation} \label{conv}
	F_\uparrow(\tau) = -\sigma_B T^4 + \int_{\tau_b}^\tau e^{-(\tau' - \tau)} \frac{d}{d\tau'} \left( \sigma_B T(\tau')^4 \right) d\tau',
\end{equation}
where $\tau_b$ is the optical depth at the bottom of the atmosphere. We assume that the upward flux at the ground is the blackbody radiation of the ground temperature ($F_{\uparrow, \mathrm{surf}} = -\sigma_B T_\mathrm{surf}^4$).
At the optical depth of the tropopause, $\tau_p$, the values of $F_\uparrow(\tau_p)$ calculated from Equations (\ref{rad}) and (\ref{conv}) should be identical to satisfy the energy balance. We search for the profiles of temperature and water vapor content that agree with energy conservation. The temperature and the water vapor mixing ratio are smoothly connected at the tropopause. 

\subsection{The initial thickness of depleted lithospheric mantle} \label{sec:degas}
The initial thickness of depleted lithospheric mantle (Figure~\ref{fig_methods}) is calculated by drawing an adiabatic temperature profile with a temperature that creates melt fraction of 0.4 at the surface. This is because the timescale for the Rayleigh-Taylor instability is predicted to be faster than the rate of solidification \citep{Maurice2017, Miyazaki2019b}, and thus the upper mantle likely has an adiabatic thermal gradient when the surface of the mantle solidifies.  The adiabatic gradient is expressed as \citep{McKenzie1984}:
\begin{equation} \label{dTdP}
	\kak{\frac{dT}{dP}}_S = \phi \frac{\alpha_l T}{\rho_l c_{p,l}} + (1-\phi) \frac{\alpha_s T}{\rho_s c_{p,s}} - \frac{T \Delta S}{c_p} \kak{\frac{d\phi}{dP}}_S,
\end{equation}
where $\alpha$ is thermal expansivity, $\rho$ is density, $c_p$ is specific heat per unit mass, and $\Delta S$ is entropy change upon melting. Subscripts $l$ and $s$ denote the melt and liquid phases, respectively. Values adopted for thermal expansivity are 3.0 and 4.6$\times$10$^{-5}$ K$^{-1}$ for $\alpha_s$ and $\alpha_l$, and, for density, $\rho_l$ and $\rho_s$ are taken to be 2900 and 3300~kg~m$^{-3}$, respectively. We adopt 1000~J~kg$^{-1}$~K$^{-1}$ for $c_p$ and 300 J kg$^{-1}$~K$^{-1}$ for $\Delta S$. Melt fraction at a given temperature and pressure is calculated using a thermodynamic model of \cite{Katz2003}, and the term $d\phi/dP$ in Equation~(\ref{dTdP}) is calculated so that changes in $T$ and $\phi$ are consistent with thermodynamics (see Supplementary of \cite{Miyazaki2022a} for details).


\end{document}